\pdfoutput=1
\documentclass{aa}  

\usepackage{graphicx}
\usepackage{txfonts}
\usepackage[version=4]{mhchem}
\newcommand{\bb}[1]{#1}

\newcommand{\bbb}[1]{#1}

\begin{document}

   \title{A new pathway to SO$_2$:}

   \subtitle{Revealing the NUV driven sulfur chemistry in hot gas giants}
    
   \author{Wiebe de Gruijter
          \inst{1},
          Shang-Min Tsai \inst{2},
          Michiel Min \inst{3},
          Rens Waters\inst{3}\fnmsep\inst{4},
          Thomas Konings \inst{5},
          Leen Decin \inst{5}
          }

   \institute{Anton Pannekoek Institute for Astronomy, University of Amsterdam,
              1090GE Amsterdam, Netherlands\\
        \and
            Department of Earth and Planetary Sciences, University of California,
        Riverside, CA, USA\\
        \and
            SRON Netherlands Institute for Space Research, Niels Bohrweg 4, NL-2333 CA Leiden, the Netherlands\\
        \and
             Department of Astrophysics/IMAPP, Radboud University Nijmegen, PO Box 9010, NL-6500 GL Nijmegen, the Netherlands\\
        \and
            Institute of Astronomy, KU Leuven, Celestijnenlaan 200D, 3001 Leuven, Belgium\\
             }

   \date{Received 3 May 2024; accepted 4 November 2024}

 
  \abstract
   {Photochemistry is a key process driving planetary atmospheres away from local thermodynamic equilibrium. Recent observations of the H$_2$ dominated atmospheres of hot gas giants have detected SO$_2$ as one of the major products of this process.}
   {We investigate which chemical pathways lead to the formation of SO$_2$ in an atmosphere, and we investigate which part of the flux from the host star is necessary to initiate SO$_2$ production.}
   {We use the publicly available S-N-C-H-O photochemical network in the VULCAN chemical kinetics code to compute the disequilibrium chemistry of an exoplanetary atmosphere.}
   {\bb{We find that there are two distinct chemical pathways that lead to the formation of SO$_2$. The formation of SO$_2$ at higher pressures is initiated by stellar flux >200 nm, whereas the formation of SO$_2$ at lower pressures is initiated by stellar flux <200 nm. In deeper layers of the atmosphere, OH is provided by the hydrogen abstraction of H$_2$O, and sulfur is provided by the photodissociation of SH and S$_2$, which leads to a positive feedback cycle that liberates sulfur from the stable H$_2$S molecule. In higher layers of the atmosphere, OH is provided by the photodissociation of H$_2$O, and sulfur can be liberated from H$_2$S by either photodissociation of SH and S$_2$, or by the hydrogen abstraction of SH.}}
   {We conclude that the stellar flux in the 200-350 nm wavelength range as well as the ratio of NUV/UV radiation are important parameters determining the observability of SO$_2$. In addition we find that there is a diversity of chemical pathways to the formation of SO$_2$. This is crucial for the interpretation of SO$_2$ detections and derived elemental abundance ratios and overall metallicities.}


\maketitle

\section{Introduction}
Giant planets on short orbital periods represent an important and well-studied group of exoplanets, \bb{due to observational biases that make it relatively easy to detect these planets and probe their atmosphere}. A major challenge currently facing the field of exoplanets is to understand their formation and evolution, so that we may place them in the wider context of the planetary systems to which they belong \citep{Boley_two_2009, Rafikov_Giant_2005}. 

The abundances of elements such as C, N, O, and S in the atmosphere provide a window into a planet's evolutionary history \citep{Turrini_Tracing_2021}. Telescopes such as Hubble and Spitzer have laid down the groundwork for characterizing the chemical composition of giant planet atmospheres \citep{Madhusudhan_Exoplanetary_2019}, and recently with the advent of the James Webb Space Telescope (JWST), our capabilities of identifying chemical species in an exoplanet's atmosphere have increased markedly. Of particular importance among the newly detected molecules is \bb{sulfur dioxide (SO$_2$)}, which has been detected by JWST in the atmospheres of WASP-39b \citep{rustamkulov_early_2023} and WASP-107b \citep{dyrek_so_2_2023}.

The presence of SO$_2$ in the atmosphere of a hot gas giant is important twofold. Firstly, the production of SO$_2$ is an outcome of photochemistry, \bb{as shown by \cite{tsai_photochemically_2023}}, which marks the first time that the effects of this physical process have been observed directly in an exoplanetary atmosphere. Secondly, its detectability depends strongly on the C/O ratio and the metallicity of the atmosphere \citep{polman_h2s_2023, dyrek_so_2_2023}, making it a tracer of two key parameters to constrain a planet's evolutionary history \citep{Oberg_effects_2011}. 

Modeling the chemistry in exoplanet atmospheres is challenging due to the need for accurate measurements of reaction rates over a wide range of pressures and temperatures to construct a chemical network \bb{\citep{chubb_data_2024}}. For this reason, \bb{thermo-}photochemical models initially focused mainly on H, C, N, and O chemistry. These models were constructed and applied to hot Jupiters by for example  \cite{Moses_Disequilibrium_2011} and \cite{Venot_chemical_2012}. A substantial effort has been undertaken in recent years to incorporate sulfur chemistry into chemical kinetics models. The first inclusion of sulfur in a chemical kinetics model of a \bb{hot} Jupiter was conducted by \cite{Zahnle_atmospheric_sulfur_2009}, who used it to study the atmospheric heating properties of hot Jupiters. In recent years, model complexity has greatly increased due to the expansion of photochemical networks and \bb{measurements of the reaction rates of additional chemical reactions}. Sulfur chemistry is now included in 1D photochemical codes such as LEVI \citep{Hobbs_sulfur_2021}, who applied it to atmospheres in the 1000-1400 K range and investigated the mixing ratios of SH, H$_2$S, and S$_2$ and their effects on the major N and C bearing molecules; VULCAN \citep{tsai_comparative_2021}, who used it to study the impact of sulfur on photochemical haze precursors; and ARGO \citep{Rimmer_Hydroxide_2021}, who applied it to the atmosphere of Venus and included clouds to explain the SO$_2$ depletion in the upper layer of the atmosphere. 

The recent observations of SO$_2$ in gas giant atmospheres and the availability of SNCHO photochemical networks call for an in-depth investigation of the chemical pathways responsible for SO$_2$ formation, and of the role of the stellar spectral energy distribution (SED). In this paper, we analyze the chemical pathways that lead to the formation of SO$_2$, building on previous work in this direction by \cite{tsai_photochemically_2023}. \bb{We expand on their work by examining which parts of the stellar SED play a part in the production of SO$_2$, and by examining its dependence on the surface gravity and atmospheric temperature.} This study provides a deeper understanding of the photochemistry occurring in the atmospheres of gas giants, which will be essential in interpreting the recent and forthcoming observations targeting hot Jupiters. 

In Section \ref{Methods}, we describe the method we use to calculate the chemical composition of the atmosphere from an initial set of elemental abundances. We also describe the method we use to compute transmission spectra from the chemical composition. In Section \ref{Fiducial Model}, we define our fiducial model\bb{, and in Section \ref{Results}, we describe our results}. In Section \ref{Discussion}, we discuss our findings and highlight the implications for observations. In Section \ref{Conclusion}, we provide a summary of our results. 

\section{Methods} \label{Methods}
\subsection{Chemical composition}
\bb{We use the photochemical kinetics code VULCAN to calculate the chemical structure of the atmosphere \citep{tsai_comparative_2021}, using a parameterized temperature-pressure (TP) profile and evolving the chemical abundances until they reach a state of equilibrium}. We use the SNCHO \bb{thermo-photochemical} network that is included in the VULCAN distribution. This network consists of 89 species\bb{, including 19 S-bearing species, 1030 thermochemical reactions (this includes forward and reverse reactions)}, and 60 photodissociation reactions. VULCAN solves the following set of Eulerian continuity equations:

\begin{equation}\label{vulcan_eqn}
    \frac{\partial n_i}{\partial t} = P_i - L_i - \frac{\partial\phi_i}{\partial z}
\end{equation}

where $n_i$ is the number density in cm$^{-3}$ of species $i$; $t$ is the time; $P_i$ and $L_i$ are the production and loss rates in cm$^{-3}$ s$^{-1}$; and $\phi_i$ is the vertical transport flux in cm$^{-2}$ s$^{-1}$. 

We use 150 vertical atmospheric layers, as is recommended for VULCAN, log-distributed from $10^3$ to $10^{-8}$ bar. For the convergence parameters we use the standard values $\delta = 0.01$\bb{, which denotes the absolute change in abundances,} and $\epsilon = 10^{-4}$, \bb{which denotes the time derivative of the abundances}, as recommended by \cite{tsai_vulcan_2017}. 

We set the elemental abundances by choosing a value for the metallicity and C/O ratio for our atmosphere. We start from the solar elemental abundances as determined by \cite{asplund_solar_abundances_2009}, and we subsequently scale the carbon abundance to obtain the required C/O ratio. We then scale the hydrogen and helium abundance to obtain the desired metallicity, where we leave the H/He ratio unchanged. We assume a constant K$_{\mathrm{zz}}$ of $10^9\,$cm$^2$/s throughout the atmosphere. 

\subsection{Temperature-pressure profile}
We use a parameterized \bb{TP} profile, which we take from the approximation derived by \cite{guillot_radiative_2010}, who modeled the atmosphere with a double-gray approximation. This approximation is described by the following equation:

\begin{equation}\label{eq:Guillot_2010}
    T^4 = \frac{3T_{\mathrm{int}}^4}{4}\left[\frac{2}{3}+\tau\right] + \frac{3T_{\mathrm{irr}}^4}{4} f \left[\frac{2}{3}+\frac{1}{\gamma\sqrt{3}}+\left(\frac{\gamma}{\sqrt{3}}-\frac{1}{\gamma\sqrt{3}}\right)\exp(-\gamma\tau\sqrt{3})\right]
\end{equation}

Here $T_{\mathrm{int}}$ is the internal temperature of the planet; $\tau$ is the optical depth to thermal radiation; $T_{\mathrm{irr}}$ is the stellar effective temperature scaled to the distance of the planet by the formula $T_{\mathrm{irr}} = T_{*}(\frac{R_*}{D})^{1/2}$, where D is the distance to the star; f scales the redistribution of heat around the planet; and $\gamma$ is the ratio of visible to thermal opacity in the atmosphere, as defined by $\gamma = \frac{\kappa_{\mathrm{v}}}{\kappa_{\mathrm{th}}}$, where $\kappa_{\mathrm{v}}$ is the atmospheric opacity to visual radiation, and $\kappa_{\mathrm{th}}$ the opacity to infrared radiation. We calculate the pressure assuming constant gravity and hydrostatic equilibrium, according to the equation $P = \frac{\tau g}{\kappa_{\mathrm{th}}}$. 

\subsection{Transmission spectrum}

To generate transmission spectra from the chemical composition and TP profile, we use ARCiS (Artful modeling Code for exoplanet Science) \citep{ormel_arcis_2019, min_arcis_2020}. Within ARCiS, we use the option to set the chemical composition of the atmosphere with the values obtained in VULCAN, and we use a spectral resolution of R=150. We include collision induced absorption and the effect of Rayleigh scattering. Opacity data is available for 27 of the 89 species included in the VULCAN SNCHO network: SO$_2$, SH, OH, O$_2$, O, NS, NO$_2$, NO, NH$_3$, NH, N$_2$O, N$_2$, HCN, H$_2$S, H$_2$O$_2$, H$_2$O, H$_2$CO, H$_2$, CS, CO$_2$, CO, CN, CH$_4$, CH, C$_2$H$_4$, C$_2$H$_2$, C$_2$. This includes all of the most abundant species, with the exception of SO, which was not included in ARCiS at the time of writing this paper. \bb{The opacity of SO is now available and deserves to be considered in future work, since SO is expected to be the parent molecule to produce SO2 \citep{tsai_photochemically_2023}. However, SO's opacity overlaps with SO$_2$ in the NIRSpec and MIRI wavelength range, making it challenging to detect this molecule.} 

ARCiS obtains line lists from the ExoMol project \citep{Tennyson_ExoMol_2016}, and the HITRAN \citep{Gordon_HITRAN_2017} and HITEMP \citep{Rothman_HITEMP_2010} databases. ARCiS uses correlated-k tables, which provide an inexpensive way to incorporate molecular lines into radiative transfer calculations \citep{lacis_description_1991, goody_correlated-k_1989}. The k-tables (at R=1000 for $\lambda = 0.3 - 50 \mu \mathrm{m}$) were taken from \cite{chubb_exomolop_2021}, who computed them using the Exocross program \citep{Yurchenko_2018}. 

\section{Fiducial model} \label{Fiducial Model}
We define a fiducial model to use as our fixed point of comparison in our investigations of SO$_2$ chemistry. In Section \ref{Results}, we will take this standard model, and we will each time vary only a single parameter to investigate its effect. 

\begin{figure}
    \centering
    \includegraphics[width=0.95\linewidth]{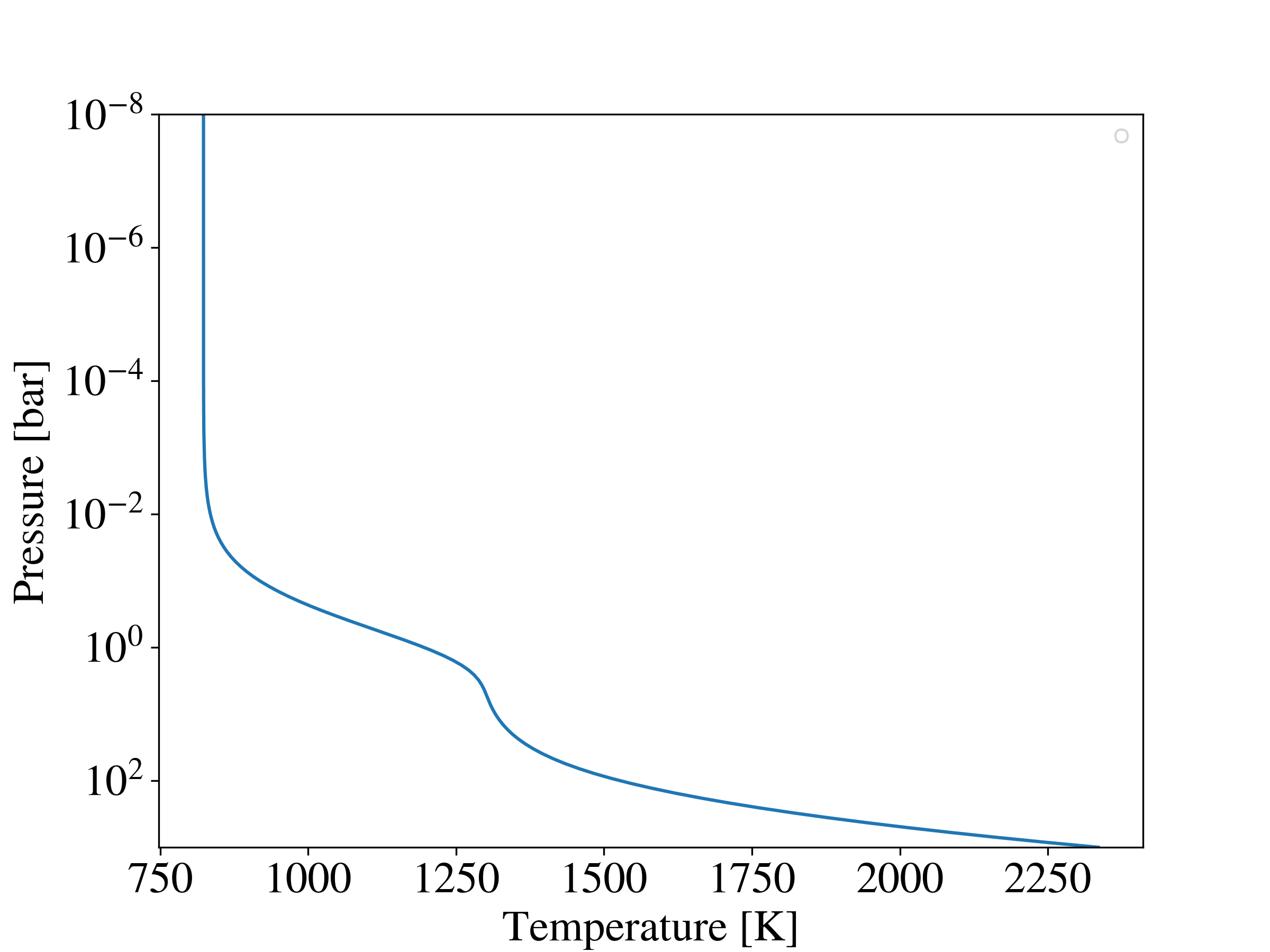}
    \caption[Temperature-Pressure profile of our fiducial model]{Temperature-pressure profile used in our fiducial model, generated using Eqn. \ref{eq:Guillot_2010}, using the parameters listed in Table \ref{tab:Atmospheric paramters for our fiducial model}.}
    \label{fig:Fiducial TP profile}
\end{figure}

We base our fiducial model on the system parameters of HD 189733 b, a \bb{well-studied} hot Jupiter around a K-type star \bb{\citep{southworth_homogeneous_2010}}. Large uncertainties still exist in the stellar radiation field of many stars due to the complications brought about by strong interstellar and telluric absorption. For \bb{HD 189733}, we \bb{use} the input stellar spectrum from \cite{Moses_Disequilibrium_2011}, who used a combination of observations of the similar star $\epsilon$ eridani and scaled solar XUV emission to reconstruct the stellar SED. We list the system parameters of \bb{HD 189733} in Table \ref{tab:HD-189733b} and the atmospheric parameters we use in Table \ref{tab:Atmospheric paramters for our fiducial model}. In Figure \ref{fig:Fiducial TP profile}, we plot the TP profile of our fiducial model. 

\bb{We choose a value for the metallicity of 10$\times$solar and a value for the C/O ratio of 0.3, based on the observed metallicities of the two known planets showing an SO$_2$ signal, WASP-107b \citep{dyrek_so_2_2023} and WASP-39b \citep{tsai_photochemically_2023}. We take the values of the parameters in Eqn. \ref{eq:Guillot_2010} from \cite{polman_h2s_2023}, who tested a range of temperature profiles to study the impact on the observability of SO$_2$. The TP profile we choose falls within the range that is expected for gas giants showing SO$_2$. It was found by \cite{polman_h2s_2023} that the temperature range over which SO$_2$ can be detected spans over a few hundred Kelvin, representing an uncertainty in our models that we explore further in Section \ref{T dependence}.}

\begin{table}[]
    \caption{:\;System parameters of \bb{HD 189733}}
    \centering
    \begin{tabular}{c|c}
        \hline
        Planet name & \bb{HD 189733 b} \\ 
        T$_*$ & 5050 K \\
        R$_*$ & 0.752 R$_{\odot}$ \\ 
        \bb{a} & 0.03142 AU \\
        R$_{\mathrm{p}}$ & 1.151 R$_{\mathrm{Jup}}$ \\
        g$_{\mathrm{p}}$ & 2250 $\mathrm{cm/s^2}$\\
        \hline
    \end{tabular}
    \label{tab:HD-189733b}
\end{table}

\begin{table}[] 
    \caption{:\; Atmospheric parameters for our fiducial model.}
    \centering
    \begin{tabular}{c|c}
        \hline
       C/O ratio  & 0.3 \\
       Metallicity  & 10x solar \\
       K$_{\mathrm{zz}}$ & $10^9$ cm$^2$/s\\
       $\gamma$ & 0.15 \\
       f & 0.1 \\
       $\kappa_{\mathrm{th}}$ & 0.01 $\mathrm{cm^2/g}$\\
       T$_{\mathrm{int}}$ & 300 K \\
       \hline
    \end{tabular}
    \label{tab:Atmospheric paramters for our fiducial model}
\end{table}

\section{Results} \label{Results}

\subsection{\bb{An alternative pathway to SO$_2$}} \label{pathway to SO2}
The production of SO$_2$ requires flux from the host star to initiate the photochemical reaction pathway. This suggests that the detectability of SO$_2$ is strongly dependent on the stellar SED. We investigate this by rescaling the UV part of the stellar input in our VULCAN models to see the effect on the mixing ratio and observability of SO$_2$. 

We take our fiducial model as defined in Section \ref{Fiducial Model}, and we remove all stellar flux below 200, 300, and 350 nm, respectively. We show the effect of this change in Figure \ref{fig:spectrum_abundances['No flux <200 nm', 'No flux <300 nm', 'No flux <350 nm']}. In this figure, we clearly see that SO$_2$ is still produced in the atmosphere and that it is  visible in the transmission spectrum even in the absence of flux below 200 and 300 nm. This is surprising, given the fact that the current SO$_2$ formation pathway suggests that the photochemical dissociation of H$_2$O is essential for the formation of SO$_2$ \citep{tsai_photochemically_2023}. 

Below we see the chemical pathway that leads to SO$_2$ production in WASP-39b, proposed by \cite{tsai_photochemically_2023}:

\begin{center}
\begin{minipage}{0.63\linewidth} 
\begin{align}
    \ce{H2O &->[h$\nu$] OH + H} \\
    \ce{H2O + H &-> OH + H2} \\
    \ce{H + H2S &-> H2 + SH} \\
    \ce{H + SH &-> H2 + S} \label{SH hydrogen abstraction} \\ 
    \ce{S + OH &-> SO + H} \\
    \ce{SO + OH &-> SO2 + H} \\ 
    \hline \nonumber \\
    \mathrm{net}: \ce{H2S + 2H2O &-> SO2 + 3H2}
\end{align}
\end{minipage}
\end{center}

\begin{figure}
    \centering
    \includegraphics[width=\linewidth]{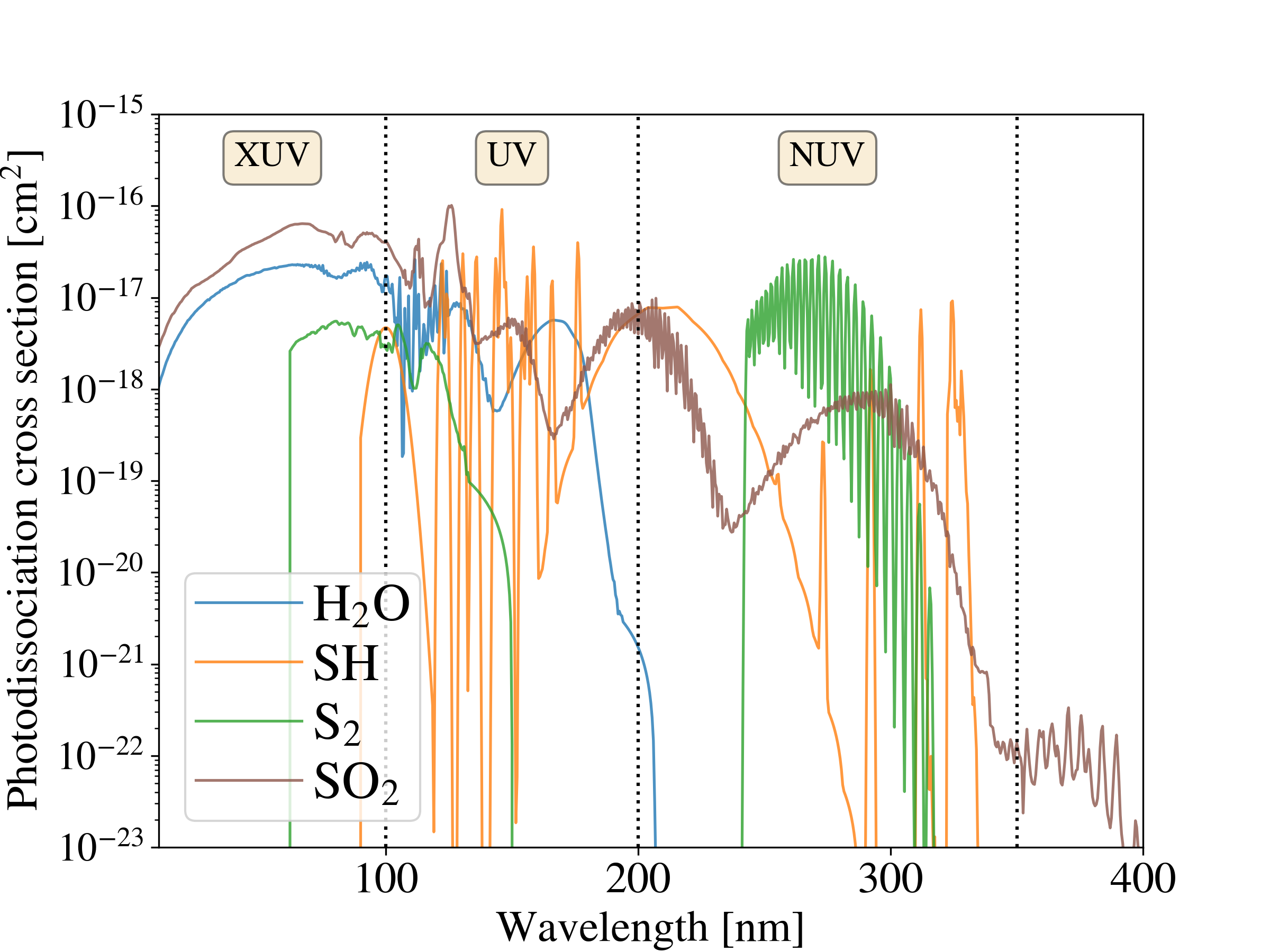}
    \caption[H2O, SH and S2 cross section]{Photodissociation cross section of H$_2$O, SH, S$_2$ and SO$_2$ as a function of wavelength. Data has been obtained from the VULCAN photochemical kinetics code, which obtains cross sections from the Leiden Observatory database \citep{Heays_photodissociation_2017}.} 
    \label{fig:H2O_SH_cross_section}
\end{figure}

We plot the photodissociation cross section of H$_2$O in Figure \ref{fig:H2O_SH_cross_section}. If H$_2$O photodissociation is a necessary condition for SO$_2$ formation, we expect the SO$_2$ signal to be completely removed if there is no flux below 300 nm, because H$_2$O will not dissociate at higher wavelengths. From our results, however, it seems sufficient to have stellar flux between 300 and 350 nm to produce SO$_2$. This suggests that the photodissociation of a different molecule is responsible for the production of SO$_2$ in this model. 

\begin{figure}
    \centering
    \includegraphics[width=\linewidth]{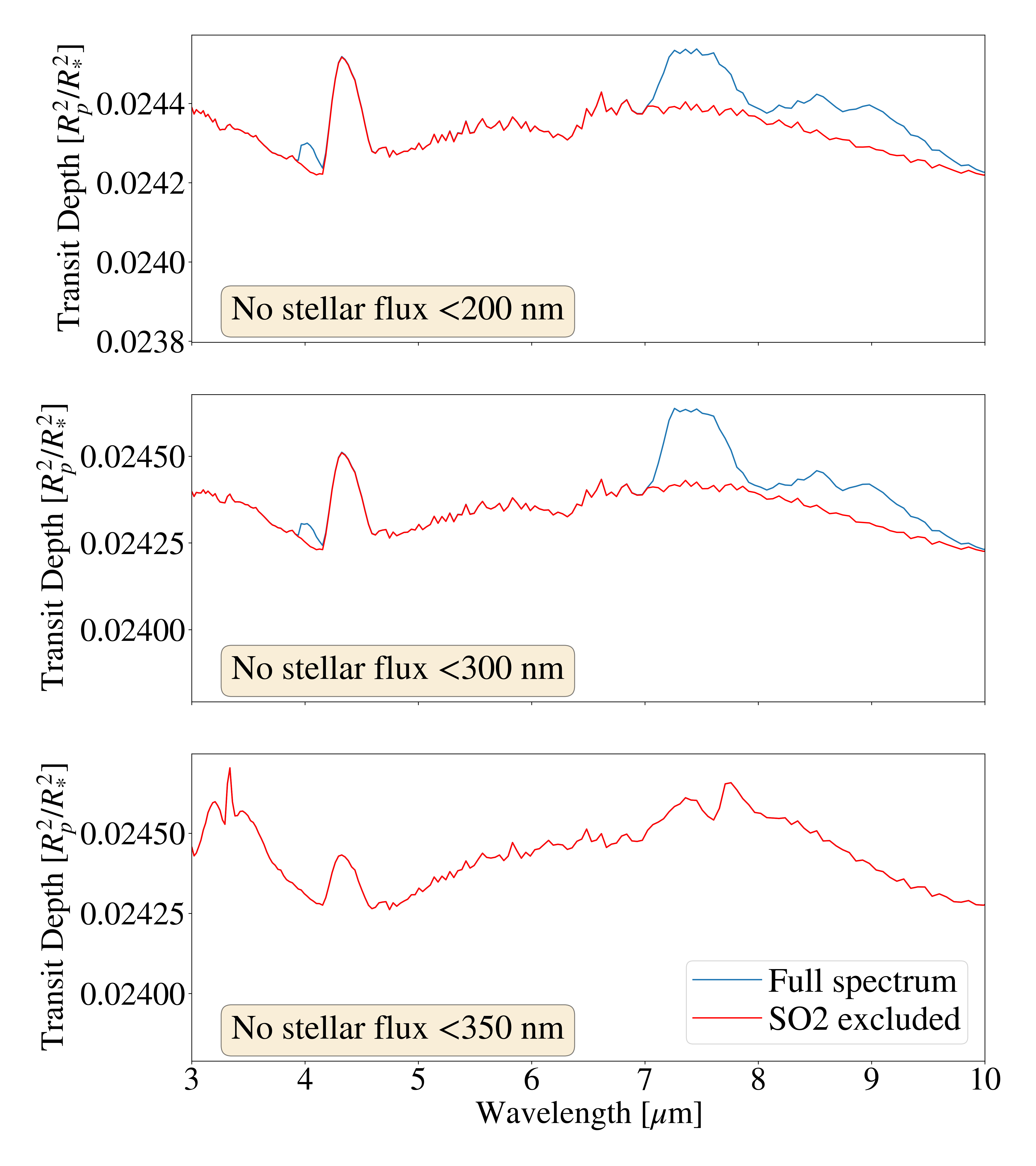}
    \caption[SO$_2$ visibility for various stellar flux levels]{\bb{Calculated transmission spectra of a VULCAN model} with all stellar flux removed $<200\,\mathrm{nm}$ (top), $<300\,\mathrm{nm}$ (middle), and $<350\,\mathrm{nm}$ (bottom). Removing the flux below 200 nm and 300 nm still produces an SO$_2$ signal, whereas removing the flux below 350 nm prevents the production of SO$_2$.}  
    \label{fig:spectrum_abundances['No flux <200 nm', 'No flux <300 nm', 'No flux <350 nm']}
\end{figure}

The approach we take here is useful, albeit limited. It provides us with key insight into the stellar fluxes that lead to SO$_2$ formation, but it is an extremely coarse approach, since removing a full part of the spectrum turns off many photochemical reactions at once. To obtain a more detailed view of the chemistry occurring in the model atmosphere, we will need a different approach. We now reanalyze our model with  the full stellar spectrum as input, and in our analysis we will turn to \bb{the effect of specific photodissociation reactions} as well as the reaction rates of specific chemical reactions to obtain more insight in the chemistry occurring in the atmosphere. 

\begin{figure}
    \centering
    \includegraphics[width=\linewidth]{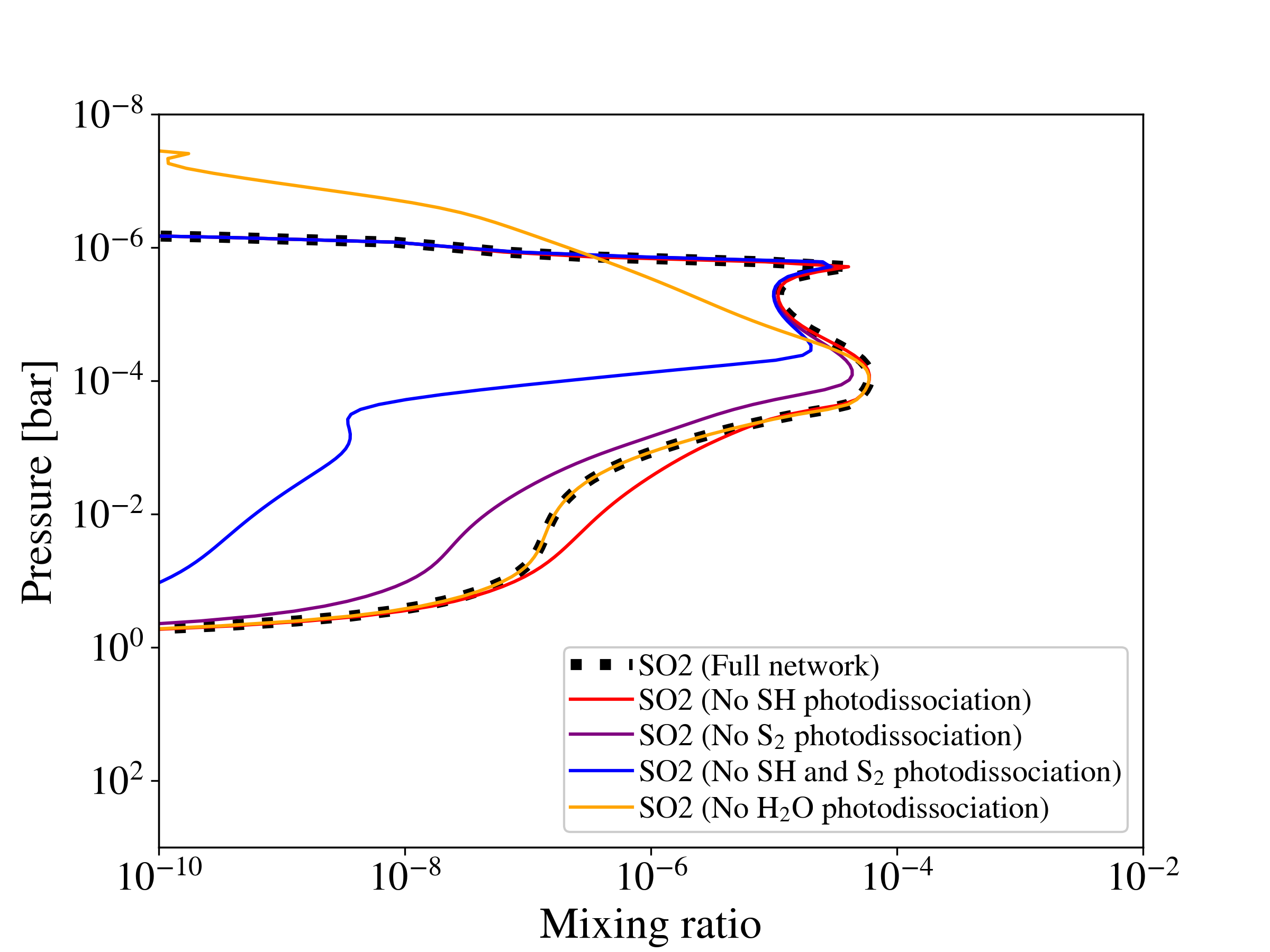}
    \caption{\bb{Mixing ratio of SO$_2$ in VULCAN models with various photodissociation reactions switched off to isolate their contribution.} If H$_2$O photodissociation is turned off, SO$_2$ is less abundant at 10$^{-6}$ bar. If both S$_2$ and SH photodissociation are turned off, SO$_2$ is less abundant at 10$^{-4}$ bar. If only S$_2$ or SH photodissociation is turned off, SO$_2$ abundance is not affected significantly.}
    \label{fig:fid_no_1121_no_1125_no_1031}
\end{figure}

\bb{In Figure \ref{fig:fid_no_1121_no_1125_no_1031}, we compare the SO$_2$ mixing ratio in our fiducial model to a model with the H$_2$O photodissociation reactions turned off. Without H$_2$O photodissociation, the mixing ratio of SO$_2$ decreases at 10$^{-6}$ bar, but at pressures larger than 10$^{-5}$ bar it remains unaltered and is still significantly larger compared to a situation without photochemistry. This supports our theory that the photodissociation of a molecule other than H$_2$O can lead to the production of SO$_2$. Based on our result from Figure \ref{fig:spectrum_abundances['No flux <200 nm', 'No flux <300 nm', 'No flux <350 nm']}, the molecule responsible needs to have significant photodissociation cross section in the 300-350\,nm wavelength range. In Figure \ref{fig:H2O_SH_cross_section}, we plot the photodissociation cross sections of SH and S$_2$. These molecules meet this criterion, suggesting their photodissociation is at the basis of an alternative chemical pathway to SO$_2$.} 


We turn our attention to the following two sets of chemical reactions: 

\begin{center}
\begin{minipage}{0.63\linewidth} 
\begin{align}
    \ce{S2 &->[h$\nu$] S + S} \label{S2 photodissociation}\\ 
    \ce{S + SH &-> H + S2} \label{S induced SH dissociation} 
\end{align}
\end{minipage}
\end{center}
and
\begin{center}
\begin{minipage}{0.63\linewidth} 
\begin{align}
    \ce{SH &->[h$\nu$] S + H} \label{SH photodissociation}\\ 
    \ce{H + H2S &-> H2 + SH} \label{H2S hydrogen abstraction}
\end{align}
\end{minipage}
\end{center}

In these reactions, we identify two \bb{chemical cycles} that eventually lead to the conversion of H$_2$S into atomic S and molecular hydrogen. The first occurs via reaction \ref{SH photodissociation} and \ref{H2S hydrogen abstraction}:

\begin{enumerate}
    \item SH is photodissociated into S and H by reaction \ref{SH photodissociation}.
    \item H reacts with H$_2$S to produce SH and H$_2$ by reaction \ref{H2S hydrogen abstraction}.
    \item \bbb{The cycle repeats from the first step:} SH is photodissociated into S and H by reaction \ref{SH photodissociation}. 
\end{enumerate}

\bb{This cycle is driven by the photochemical dissociation of SH, which leads to the production of atomic S and H. Atomic S leaves the cycle, while atomic H reacts with H$_2$S to produce SH, which brings us back to the start of the cycle. This leads to the rapid conversion of the stable H$_2$S into atomic S and molecular H$_2$.} An alternative feedback loop is slightly more complicated but leads to the same outcome, and occurs via reaction \ref{S2 photodissociation}, \ref{S induced SH dissociation}, and \ref{H2S hydrogen abstraction}:

\begin{enumerate}
    \item \bbb{S$_2$} is photodissociated into S by reaction \ref{S2 photodissociation}.
    \item SH reacts with S to produce H and S$_2$ by reaction \ref{S induced SH dissociation}.
    \item H reacts with H$_2$S to produce SH and H$_2$ by reaction \ref{H2S hydrogen abstraction}.
    \item \bbb{The cycle repeats from the first step: S$_2$} is photodissociated into S by reaction \ref{S2 photodissociation}.
\end{enumerate}

\bb{Both of these pathways dissociate SH into H and S. Atomic H reacts with H$_2$S to produce SH, which is dissociated again by the effects of the stellar radiation field.} In this way, photochemistry is \bb{indirectly} responsible for the liberation of \bb{atomic S} from the H$_2$S molecule.

\bb{This provides a source of atomic S. However, to produce SO$_2$, we also require two atoms of oxygen. These are provided via a pathway that is mechanistically similar to the one laid out above:}

\begin{center}
\begin{minipage}{0.63\linewidth} 
\begin{align}
    \ce{SH &->[h$\nu$] S + H} \\
    \ce{H2 + S &-> SH + H}  \\
    \ce{H2O + H &-> OH + H2} 
\end{align}
\end{minipage}
\end{center}

\bb{The photodissociation of SH produces atomic S, which reacts with molecular H$_2$ to produce SH. This cycle converts a fraction of the available molecular H$_2$ into atomic H. This leads to an increased availability of atomic H, which reacts with H$_2$O to produce OH. This provides all the ingredients necessary to produce SO$_2$. Atomic S reacts with OH to produce SO, and SO again reacts with OH to produce SO$_2$.}

\begin{center}
\begin{minipage}{0.63\linewidth} 
\begin{align}
    \ce{S + OH &-> SO + H} \\
    \ce{SO + OH &-> SO2 + H} 
\end{align}
\end{minipage}
\end{center}

\bb{In Figure \ref{fig:fid_no_1121_no_1125_no_1031}, we turn off the photodissociation of either SH or S$_2$, as well as both of them at the same time. We see that turning both of them off significantly reduces the mixing ratio of SO$_2$ around 10$^{-4}$ bar. Turning off only one of them does not reduce the SO$_2$ mixing ratio significantly. This leads us to conclude that either one of these reactions is sufficient to lead to the production of SO$_2$ at 10$^{-4}$ bar.}



\subsection{\bb{Rate analysis}}

\begin{figure}
    \centering
    \includegraphics[width=\linewidth]{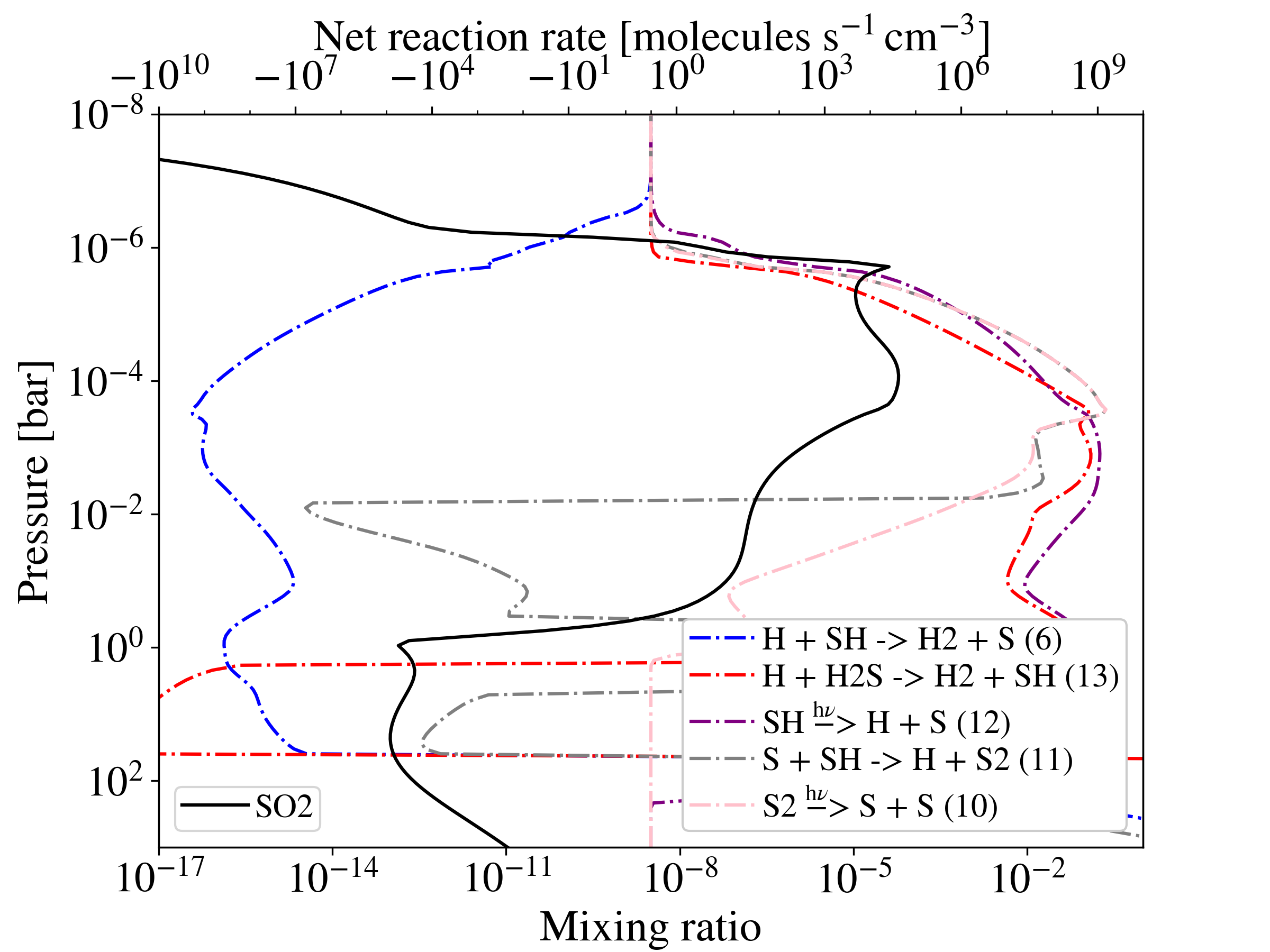}
    \caption[H2S Positive Feedback Loop]{\bb{Mixing ratio of SO$_2$ (solid line) and net reaction rates of various chemical reactions, subtracting the reverse reaction rate from the forward reaction rate (dash-dot lines) in the fiducial model}. We see that reactions \ref{S2 photodissociation}, \ref{S induced SH dissociation}, \ref{SH photodissociation}, and \ref{H2S hydrogen abstraction} are particularly pronounced in \bb{the} region of the atmosphere where SO$_2$ is abundant.}
    \label{fig:Rates_density[1121]['SO2']}
\end{figure}

\bb{In this section, we turn to an analysis of the reaction rates of the chemical reactions discussed in the last section, to investigate which specific reactions are active in the various scenarios discussed above. In Figure \ref{fig:Rates_density[1121]['SO2']} we plot the reaction rates of reactions \ref{SH hydrogen abstraction}, \ref{S2 photodissociation}, \ref{S induced SH dissociation}, \ref{SH photodissociation} and \ref{H2S hydrogen abstraction} in our fiducial model. We see that these reactions are active at similar rates in the layers where SO$_2$ is abundant, providing a further indication for their involvement in the pathway to SO$_2$. Further analysis shows that the model reaches near-chemical equilibrium in the deeper parts of the atmosphere (see Figure \ref{fig:Normalized_rates}).}

\begin{figure}
    \centering
    \includegraphics[width=\linewidth]{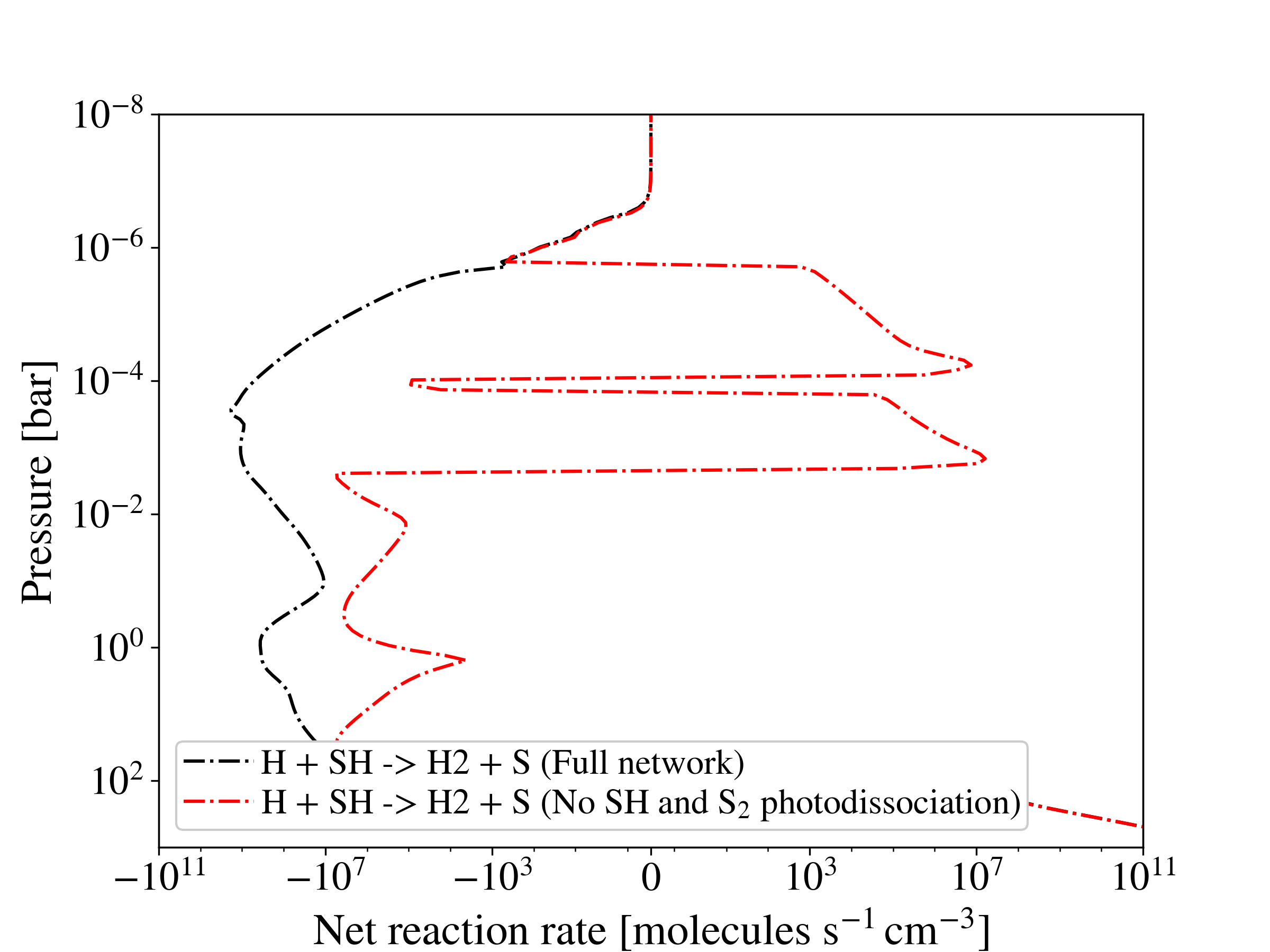}
    \caption[Source of S in absence of SH and S$_2$ photodissociation]{\bb{Net reaction rate of SH hydrogen abstraction, subtracting the reverse reaction rate from the forward reaction rate}. This figure shows that in the absence of SH and S$_2$ photodissociation, at 10$^{-6}$ bar, hydrogen abstraction of SH is able to take over as the source of atomic S in the pathway to SO$_2$.}
    \label{fig:fid_no_1121_1125_Rates_[611]}
\end{figure}

\bb{From Figure \ref{fig:fid_no_1121_no_1125_no_1031}, it became clear that around 10$^{-6}$ bar, the mixing ratio of SO$_2$ is unaltered by the absence of SH and S$_2$ photodissociation. This raises the question what the source of atomic sulfur is at this layer without these reactions. In Figure \ref{fig:fid_no_1121_1125_Rates_[611]}, we see that the hydrogen abstraction of SH takes over as the source of atomic S at 10$^{-6}$ bar. The photodissociation of H$_2$O at these layers provides sufficient amounts of atomic H to drive this reaction forward. In deeper layers of the atmosphere H$_2$O photodissociation does not take place, and there is insufficient production of atomic H for the hydrogen abstraction of SH to occur, preventing the formation of atomic S.}

\bb{In addition, Figure \ref{fig:fid_no_1121_no_1125_no_1031} showed that turning off H$_2$O photodissociation reduces the \bb{mixing ratio} of SO$_2$ around 10$^{-6}$ bar, but not at higher pressures around 10$^{-4}$ bar. In Section \ref{pathway to SO2}, we claimed it was the hydrogen abstraction of H$_2$, driven by the photodissociation of SH, that drove the production of atomic H, which subsequently led to the hydrogen abstraction of H$_2$O. We investigate this in Figure \ref{fig:Rates_density[1, 1031]['SO2']},  where we plot the reaction rates of the photodissociation and hydrogen abstraction of H$_2$O. At lower pressures, OH is provided by photodissociation of H$_2$O, but at higher pressures, it is provided by the hydrogen abstraction of H$_2$O. This confirms our claim that there are two separate pathways to SO$_2$, depending on the layer of the atmosphere.}

\begin{figure}
    \centering
    \includegraphics[width=\linewidth]{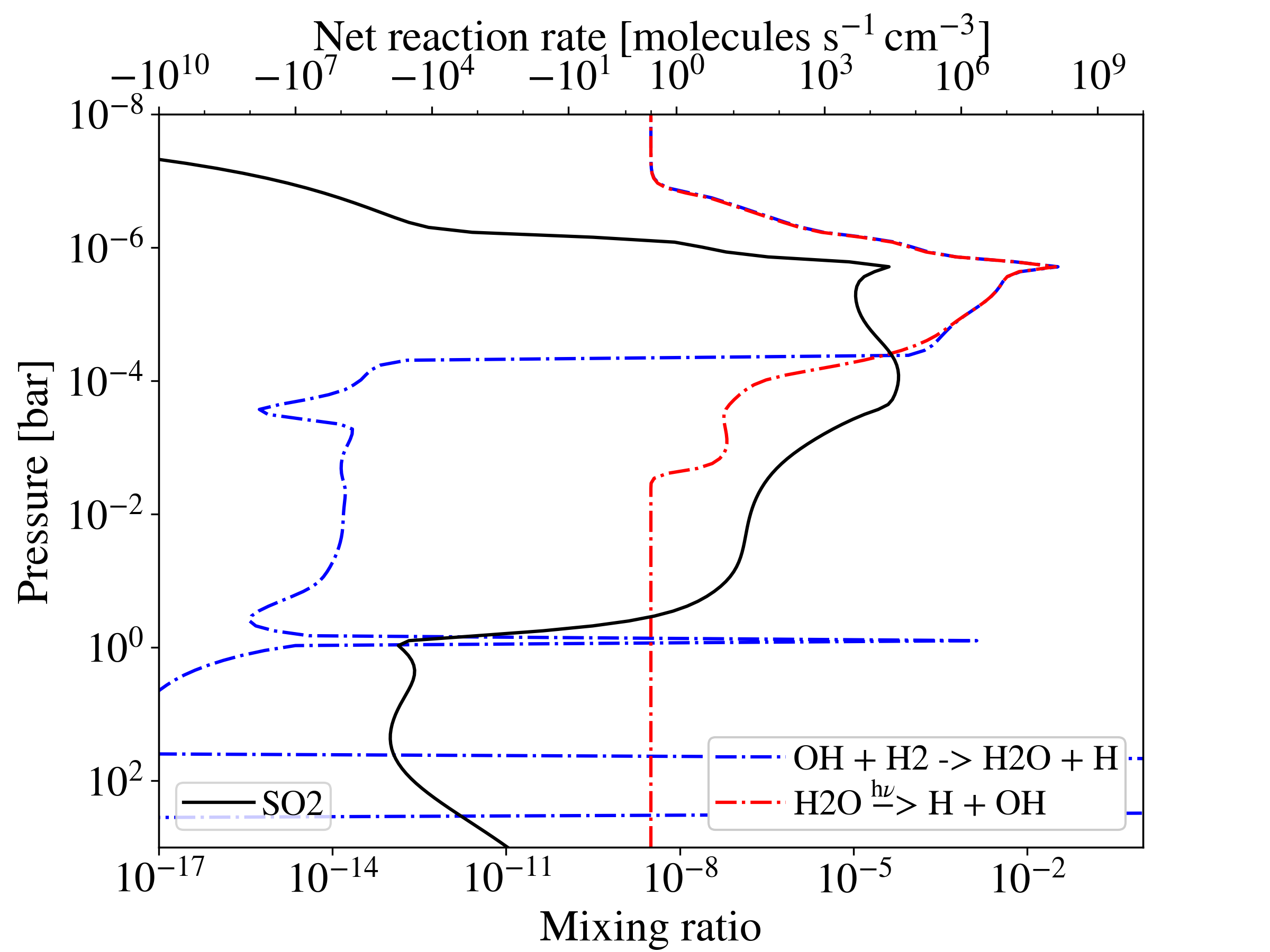}
    \caption[Reaction rates of photodissociation and hydrogen abstraction of H$_2$O]{\bb{Mixing ratio of SO$_2$ (solid line) and net reaction rates of H$_2$O hydrogen abstraction and photodissociation, subtracting the reverse reaction rate from the forward reaction rate (dash-dot lines) in the fiducial model}. At $10^{-6}$ bar, OH is supplied by H$_2$O photodissociation. At $10^{-4}$ bar, OH is supplied by H$_2$O hydrogen abstraction.}
    \label{fig:Rates_density[1, 1031]['SO2']}
\end{figure}

\bb{This leaves us with an unresolved problem: the mechanism responsible for producing SO$_2$ in our model with no flux below 300 nm. In Figure \ref{fig:H2O_SH_cross_section}, we plot the photodissociation cross section of SH and S$_2$ as a function of wavelength. We see that there is still significant absorption in the range 300-350 nm, which is in line with our result that stellar flux in this wavelength range is sufficient to lead to the production of SO$_2$. Figure \ref{fig:10solar_CO0.3_fluxrescale_2_300_0/Rates_density[1, 1031]['SO2']} shows the reaction rate of H$_2$O hydrogen abstraction, which shows that without flux at wavelengths shorter than 300$\,$nm this reaction shifts completely to the right. This suggests that in a model atmosphere without flux at wavelengths shorter than 300$\,$nm, hydrogen abstraction of H$_2$O completely takes over as the source of OH in the pathway to SO$_2$.}

\bb{The overarching scenario that arises is that the SH-H$_2$S feedback loop is the main provider of atomic S for the production of SO$_2$. H$_2$O photodissociation is the main provider of OH at pressures around $10^{-6}$ bar, whereas H$_2$O hydrogen abstraction is responsible for providing OH at higher pressures of approximately $10^{-4}$ bar.}

\begin{figure}
    \centering
    \includegraphics[width=\linewidth]{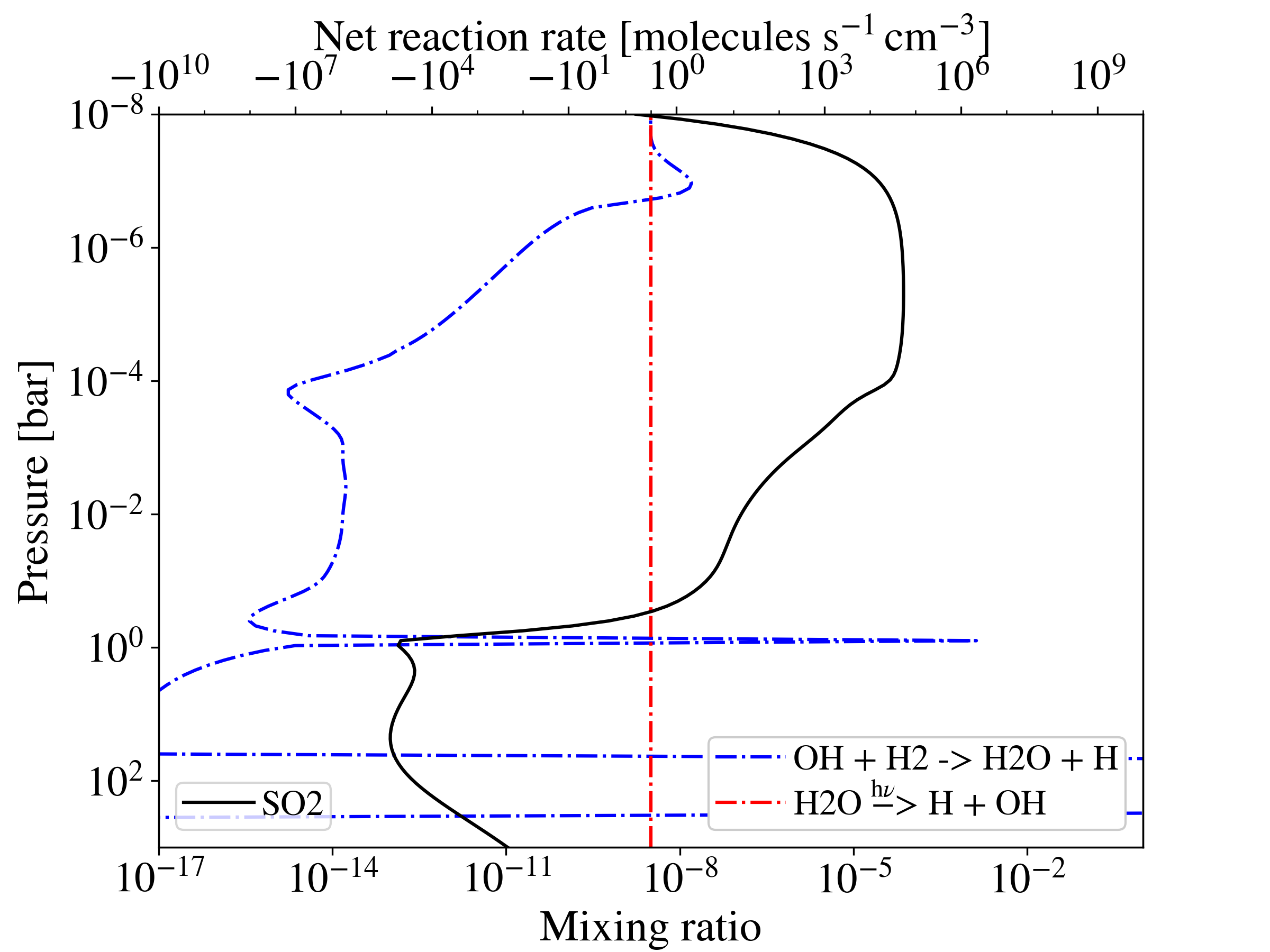}
    \caption[Hydrogen abstraction of water in the case of nu UV flux < 300 nm]{\bb{Mixing ratio of SO$_2$ (solid line) and net reaction rates of H$_2$O hydrogen abstraction and photodissociation, subtracting the reverse reaction rate from the forward reaction rate (dash-dot lines) in the model with no flux <300\,nm}. We see that \bb{in a model without stellar flux <300\,nm,} no H$_2$O photodissociation occurs, and that OH is produced by H$_2$O hydrogen abstraction in almost all layers of the atmosphere.}
    \label{fig:10solar_CO0.3_fluxrescale_2_300_0/Rates_density[1, 1031]['SO2']}
\end{figure}

\begin{figure}
    \centering
    \includegraphics[width=\linewidth]{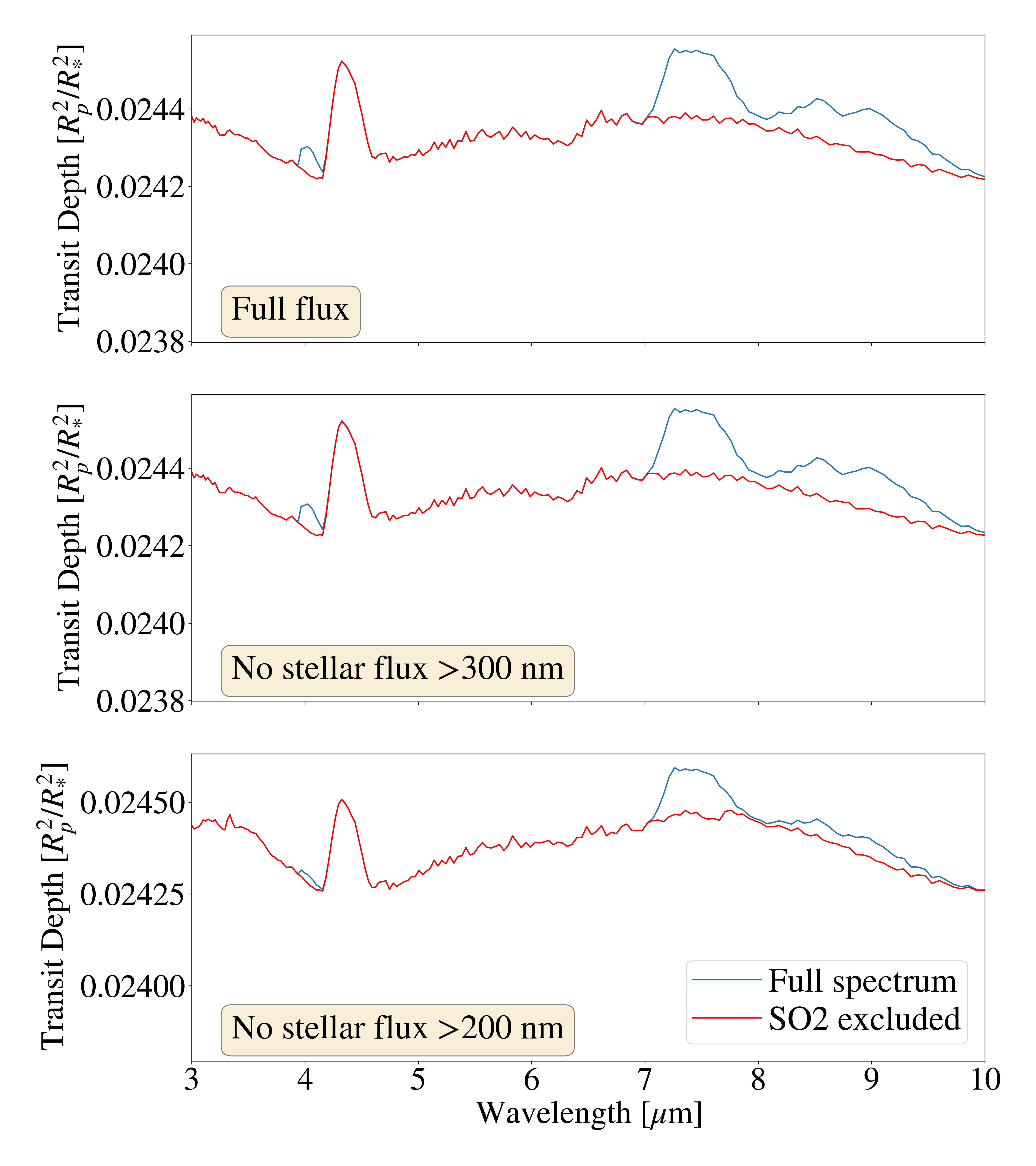}
    \caption{\bb{The transit spectrum of} our fiducial model (top panel), the same model but with all flux removed above 300 nm (middle panel) and all flux removed above 200 nm (bottom panel). This figure shows the contribution of the flux between 200 and 300 nm and the contribution of flux >300 nm to the transit spectrum.}
    \label{fig:fid_0_200_800_0_300_800_spectrum_abundances}
\end{figure}

In a final experiment on the stellar flux, we run models with all flux removed above 200 and 300 nm to quantify the impact of these parts of the SED on the transit spectrum. We show these results in Figure \ref{fig:fid_0_200_800_0_300_800_spectrum_abundances}. Here we see that flux between 200 and 300 nm contributes significantly to the visibility of SO$_2$ in the transit spectrum, whereas flux at wavelengths above 300 nm does not. However, we saw in Figure \ref{fig:spectrum_abundances['No flux <200 nm', 'No flux <300 nm', 'No flux <350 nm']} that flux between 300 and 350 nm produces a significant amount of SO$_2$ in the absence of flux below 300 nm. We conclude that flux between 300 and 350 nm only contributes significantly to the formation of SO$_2$ when no flux at lower wavelengths is present that photodissociates SO$_2$. This situation occurs in deeper parts of the atmosphere, which are shielded from XUV and UV radiation that dissociates SO$_2$, while it is still subjected to NUV radiation above 300\,nm that leads to the production of SO$_2$. This effect is responsible for the dependence of SO$_2$ formation on surface gravity, which we show in Section \ref{surface_gravity}. 

\subsection{The influence of surface gravity} \label{surface_gravity}
Inspired by the recent detection of SO$_2$ in the atmosphere of WASP-107b \citep{dyrek_so_2_2023}, a planet with an extraordinarily low surface gravity \bb{of 309\,cm/s$^2$}, we investigate the influence of the surface gravity on the mixing ratios and visibility of SO$_2$. To do this, first we change the TP profile of our fiducial model to an isothermal TP profile. The reason we choose to do this is to allow us to isolate the effect of surface gravity. The parameterized TP profile we use, as described in Section \ref{Methods}, is a parameterization that depends on the surface gravity, meaning that if we change the surface gravity we also change the TP profile. Replacing the TP profile with an isothermal profile does not significantly impact the \bb{mixing ratios} of SO$_2$, likely due to the fact that our parameterized TP profile is also isothermal below a pressure of 10$^{-2}$ bar. We set the isothermal temperature of our atmosphere at T = 740$\,$K, the effective temperature of WASP-107b \bb{\citep{dyrek_so_2_2023}}. 

\begin{figure}
    \centering
    \includegraphics[width=\linewidth]{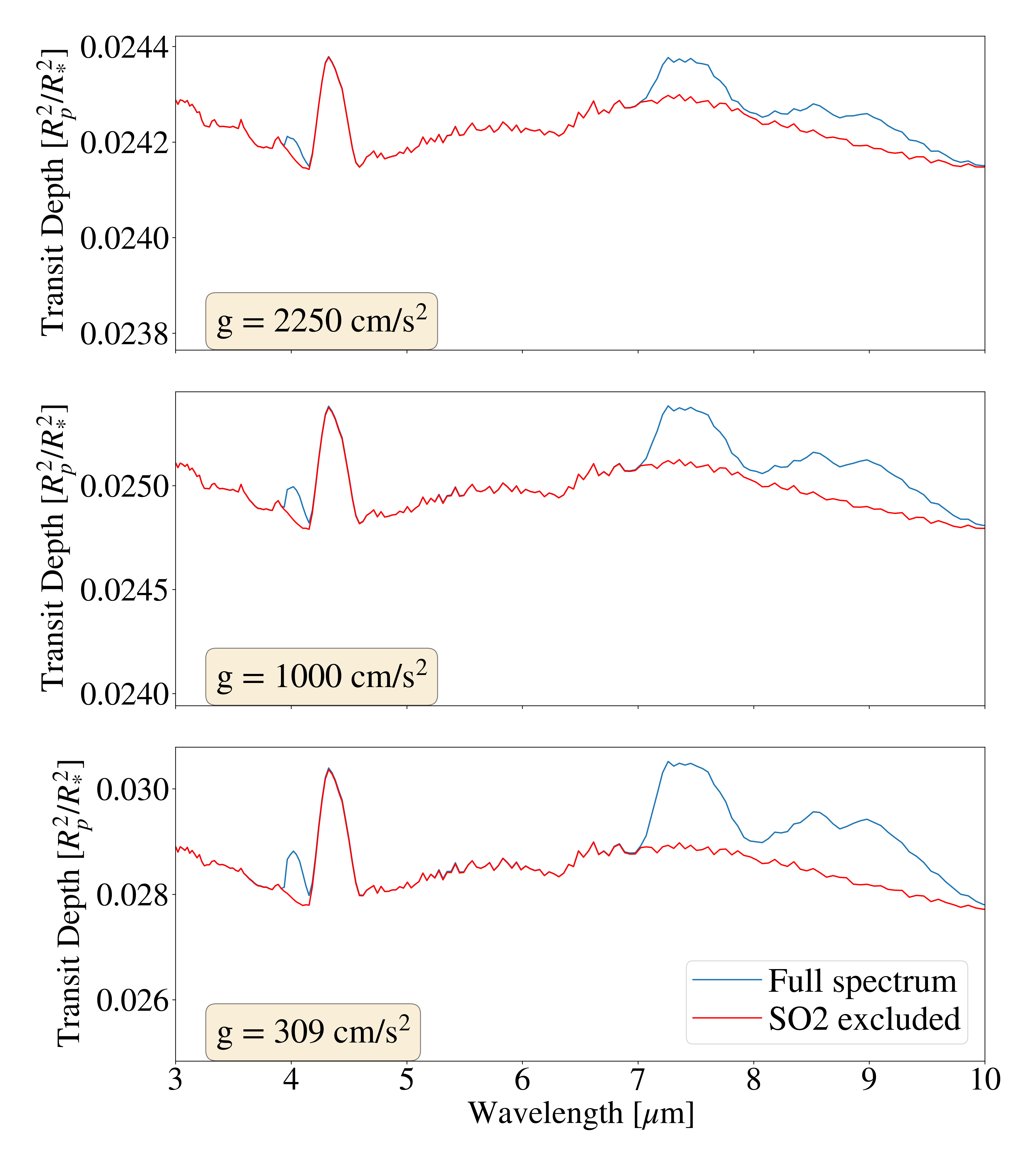}
    \caption{The visibility of SO$_2$ in the transmission spectrum as a function of surface gravity in the model with an isothermal TP profile of 740 $\,$K. We see that a lower surface gravity leads to a larger visibility of SO$_2$ in the transmission spectrum.}
    \label{fig:g309_1000_gp_spectrum_abundances_hide_SO2}
\end{figure}

\begin{figure}
    \centering
    \includegraphics[width=\linewidth]{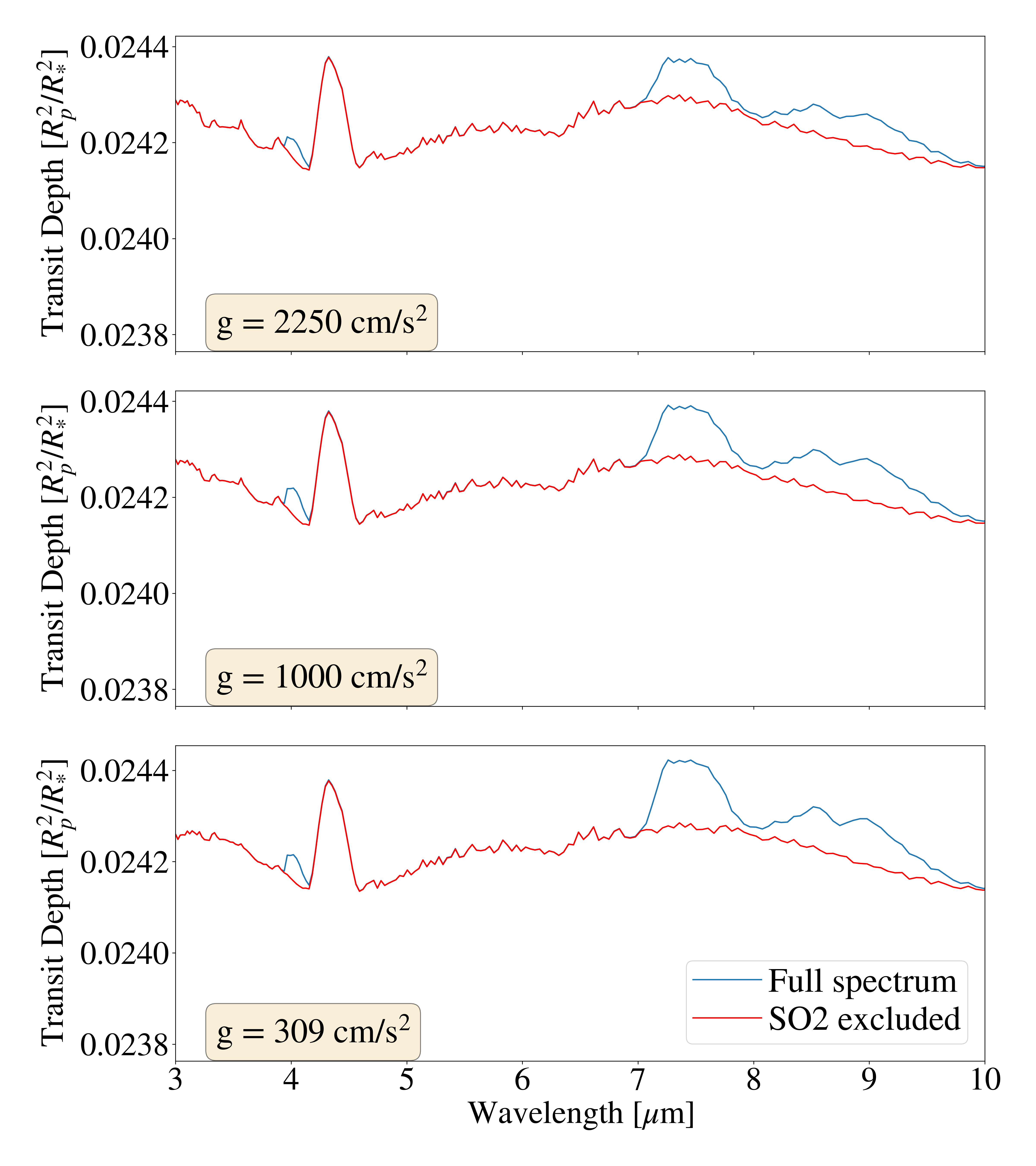}
    \caption{The same as \bb{Figure} \ref{fig:g309_1000_gp_spectrum_abundances_hide_SO2}, but with the surface gravity set to 2250 cm/s$^2$ for all models when computing the spectrum, so that the change in the spectrum is attributable only to the change in mixing ratios.}
    \label{fig:g309_1000_gp_spectrum_abundances_hide_SO2_gspectrump}
\end{figure}

In Figure \ref{fig:g309_1000_gp_spectrum_abundances_hide_SO2}, we plot the transmission spectra of our isothermal model for three values of the surface gravity. We see in this plot that the visibility of SO$_2$ in the transmission spectrum increases as the surface gravity decreases. In computing the spectrum, the pressure coordinates are translated to geometric coordinates under the assumption of hydrostatic equilibrium. A lower surface gravity translates to an extended scale height and therefore increased transit depth, which causes part of the increase in visibility of SO$_2$. Therefore, to isolate the effect of the increased mixing ratios on the visibility of SO$_2$ in the spectrum, in Figure \ref{fig:g309_1000_gp_spectrum_abundances_hide_SO2_gspectrump}, we take the mixing ratios of these atmospheres but keep the surface gravity at 2250 cm/s$^2$ when computing the transmission spectrum. This produces a similar result although the effect is less pronounced, from which we conclude that the effect on the transmission spectrum is due to both the larger scale height as well as the increase in SO$_2$ mixing ratio brought about by lower surface gravity. 

\begin{figure}
    \centering
    \includegraphics[width=\linewidth]{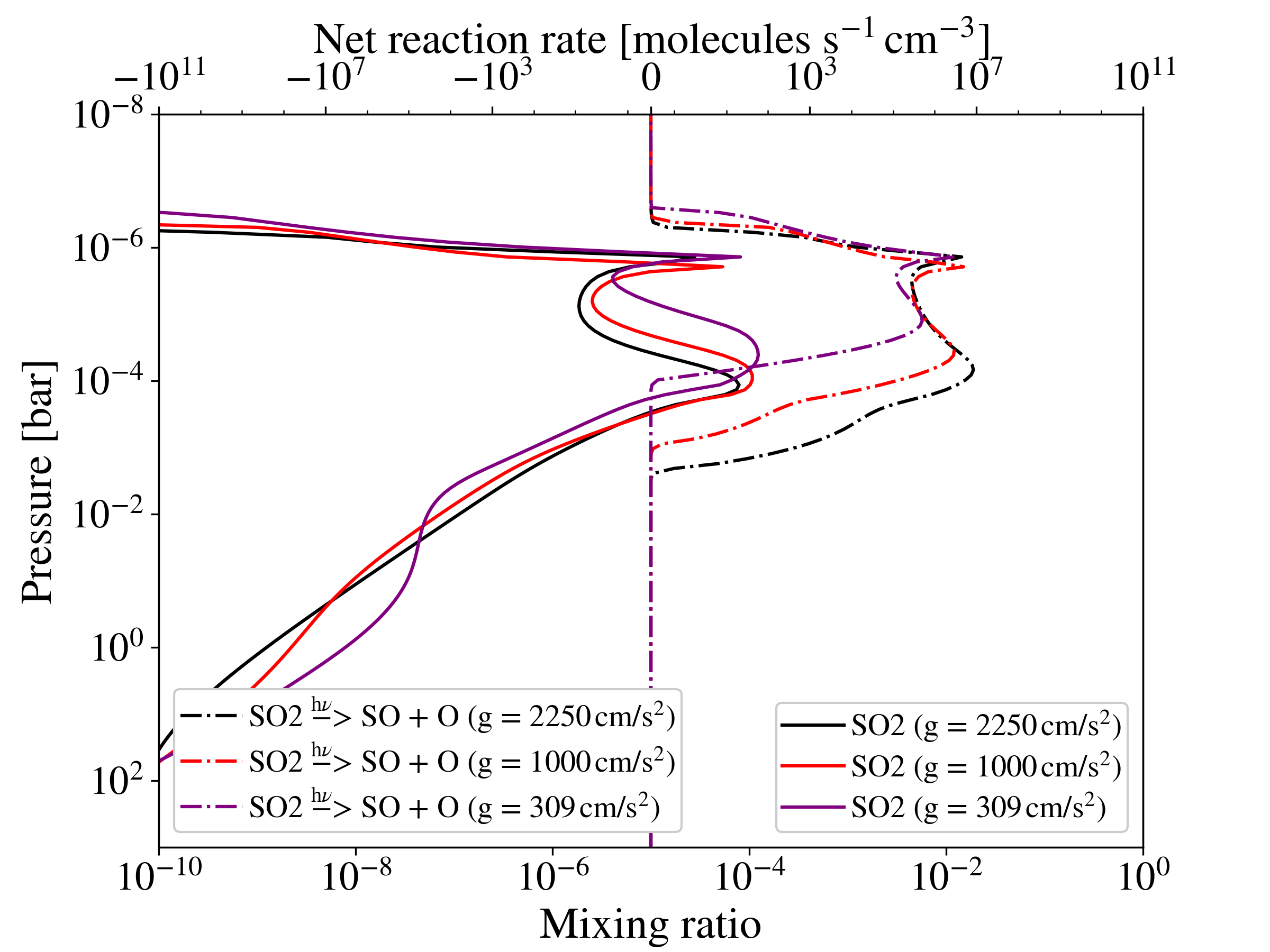}
    \caption{The \bb{mixing ratios} of SO$_2$ (solid lines) and the rate of the dominant SO$_2$ photodissociation branch (dash-dot lines) in the fiducial model with an isothermal TP profile of 740 K. We see that the \bb{mixing ratio} of SO$_2$ is elevated and photodissociation of SO$_2$ is diminished in the models with lower surface gravity.}
    \label{fig:g_309_g1000_gp_1117_SO2}
\end{figure}

In Figure \ref{fig:g_309_g1000_gp_1117_SO2}, we plot the mixing ratios of SO$_2$ in these three models, as well as the rate of SO$_2$ photodissociation. We see that the mixing ratio of SO$_2$ increases and that the photodissociation rate of SO$_2$ decreases as the surface gravity is decreased. We interpret this as the stellar UV radiation reaching higher pressure parts of the atmosphere in a model with higher surface gravity, which can be expected from the relation between optical depth, pressure, and surface gravity (assuming constant gravity and hydrostatic equilibrium):

\begin{centering}
\begin{equation}
    P = \frac{\tau g}{\kappa_{\mathrm{th}}}
\end{equation}
\end{centering}

Here P is the pressure, $\tau$ is the optical depth, g is the surface gravity, and $\kappa_{\mathrm{th}}$ is the opacity to infrared radiation. At lower surface gravity, the point where $\tau = 1$ is reached for radiation at a particular wavelength is at a lower pressure. We conclude that lower surface gravity prevents SO$_2$ photodissociation from reaching the deeper layers of the atmosphere. The NUV flux that produces SO$_2$ at these layers is also absorbed in higher layers, so the natural question that arises is why this nonetheless leads to an increase in mixing ratios. 

In Figure \ref{fig:spectrum_abundances['No flux <200 nm', 'No flux <300 nm', 'No flux <350 nm']}, we showed that in a situation with only stellar flux above 300 nm there was a strong production of SO$_2$. However, Figure \ref{fig:fid_0_200_800_0_300_800_spectrum_abundances} showed that removing the flux above 300 nm did not impact the production of SO$_2$. We concluded that stellar flux between 300 and 350 nm only produces SO$_2$ in the absence of stellar flux at shorter wavelengths that dissociates SO$_2$. 

\begin{figure}
    \centering
    \includegraphics[width=\linewidth]{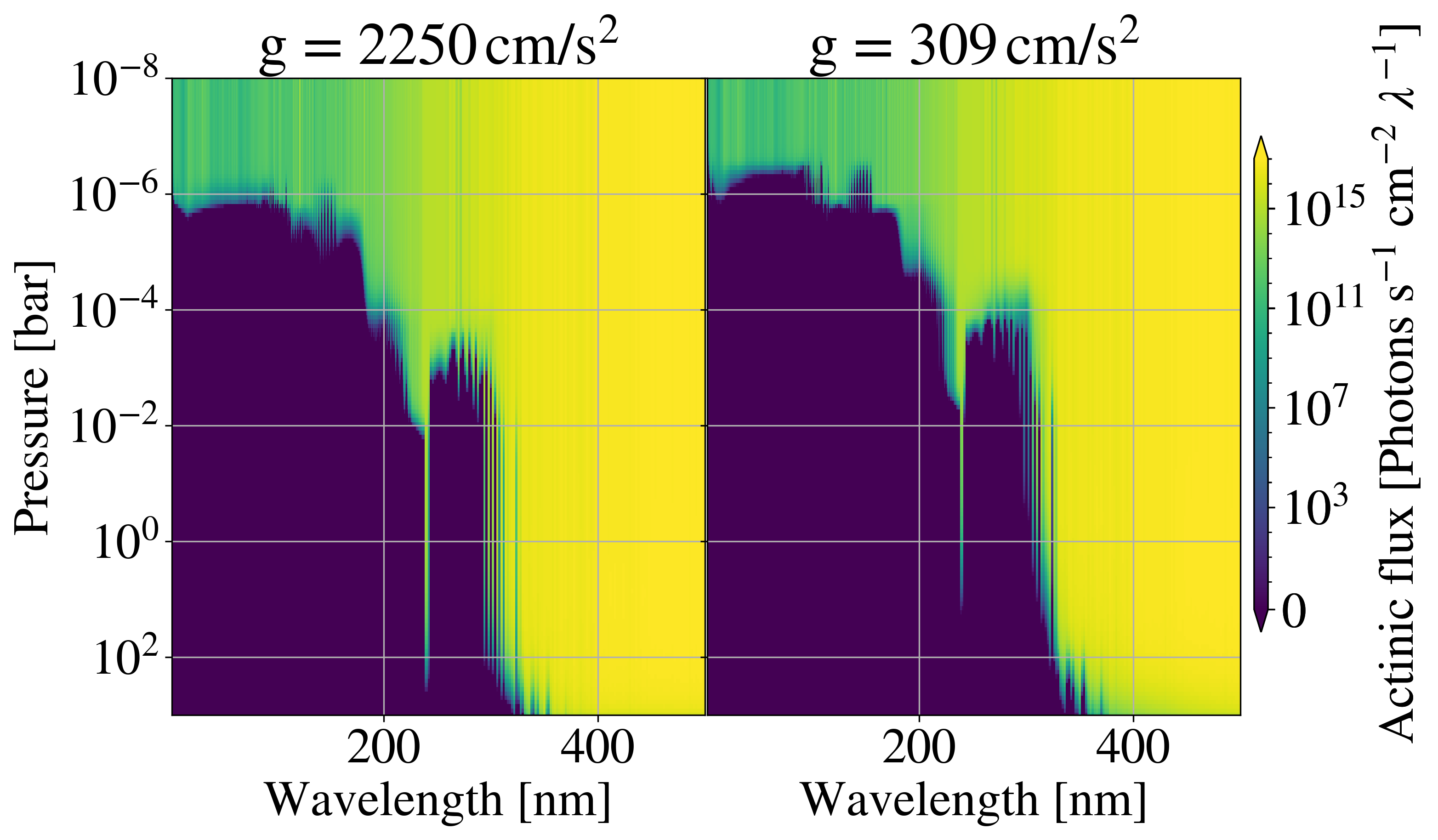}
    \caption{Actinic flux in the atmosphere for two values of the surface gravity. Stellar flux is absorbed in higher layers of the atmosphere for a lower surface gravity.}
    \label{fig:g309_gp_actinic_flux}
\end{figure}

A situation similar to this arises deep in an atmosphere with a low surface gravity. To show this, in Figure \ref{fig:g309_gp_actinic_flux} we plot the actinic flux (the radiation intensity integrated over all directions) in an atmosphere with high and low surface gravity. In both situations, stellar flux between 300 and 350 nm reaches deeply into the atmosphere. At pressures higher than 10$^{-4}$ bar, this does not lead to the production of significant amounts of SO$_2$, because high pressures strongly drive the abundances towards chemical equilibrium. Low surface gravity prevents stellar flux shorter than 200 nm from reaching the layers around 10$^{-4}$ bar. The photodissociation cross section of SO$_2$ goes down as a function of wavelength (see Figure \ref{fig:H2O_SH_cross_section}), so while flux in the 300-350 nm window is still able to reach the layers around 10$^{-4}$ bar when the surface gravity is low, which leads to the production of SO$_2$, the photodissociation of SO$_2$ is strongly diminished. This leads to an increase in the mixing ratio of SO$_2$, which in turn leads to an increased visibility in the transit spectrum. We do note that the SO$_2$ peak shifts upwards slightly, so the presence of flux between 300 and 350 nm is not completely able to compensate for the upward shift of the NUV absorption in the atmosphere. 

\subsection{The influence of temperature}\label{T dependence}
Reaction rates in an atmosphere are strongly dependent on the temperature of the atmosphere, and uncertainties still exist in the TP structure of exoplanetary atmospheres. For these two reasons, we investigate the effect of changing the temperature of our model atmosphere on the chemical pathway to SO$_2$. We change the temperature by shifting it throughout the vertical column by a set amount. 

\begin{figure}
    \centering
    \includegraphics[width=\linewidth]{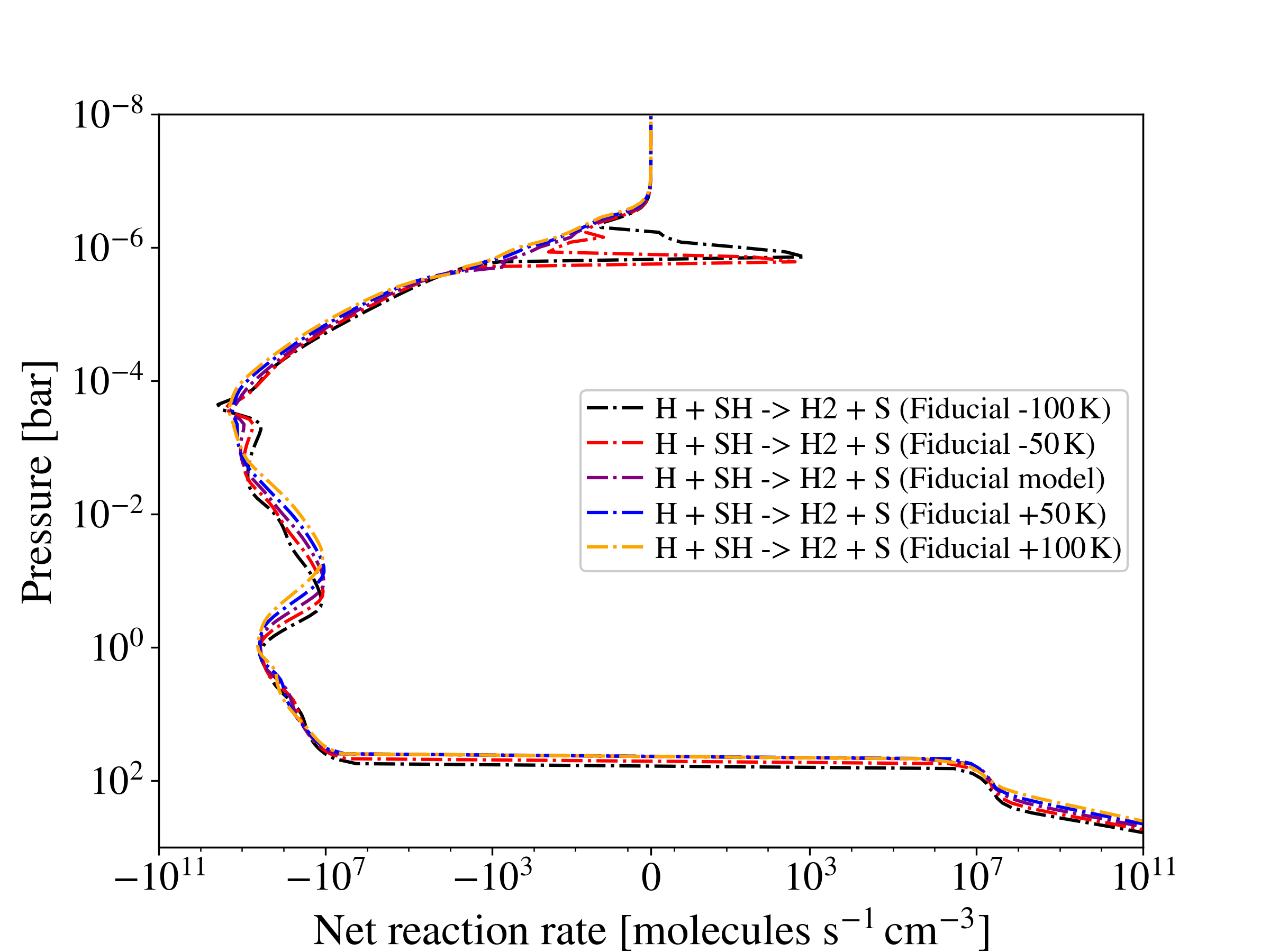}
    \caption{\bb{The net reaction rate of SH hydrogen abstraction, subtracting the reverse from the forward rate. We show the output of five VULCAN models, one with the fiducial TP profile as described in Figure \ref{fig:Fiducial TP profile}, the others with the temperature profile shifted by a set amount throughout the atmospheric column.} We see that if we decrease the temperature, at a pressure of 10$^{-6}$ bar, SH reacts with atomic H to produce atomic sulfur.} 
    \label{fig:T_dependence_SH_dissociation}
\end{figure}

We investigate the effect of changing the temperature of our atmosphere by 100$\,$K and 50$\,$K. The only significant difference we find is the direction of the reaction \ce{SH + H -> S + H2} in the upper layer of the atmosphere. We show this effect in Figure \ref{fig:T_dependence_SH_dissociation}. Here we see that this reaction runs to the right at $10^{-6}$ bar if we lower the temperature of the atmosphere, whereas if we increase it, it still runs to the left. 

This suggests a temperature dependence of the pathway to SO$_2$ in the upper layer of the atmosphere. At lower temperatures below $\sim800\,$K, around $10^{-6}$ bar, the pathway to SO$_2$ does occur as described by \cite{tsai_photochemically_2023}. Around $10^{-4}$ bar, photodissociation of SH and S$_2$ is still the pathway to SO$_2$, even at lower temperatures. At higher temperatures, the reaction \ce{SH + H -> S + H_2} runs to the left, and photochemical dissociation of SH and S$_2$ drives the liberation of sulfur from H$_2$S at all layers. 


\section{Discussion} \label{Discussion}
We have shown that stellar flux between 300 and 350 nm is a sufficient condition for the formation of SO$_2$. Furthermore, the photodissociation of SH and S$_2$ plays a significant role in the formation of SO$_2$ in deeper layers of the atmosphere. Hydrogen abstraction of H$_2$O is the predominant source of OH in these deeper layers, and photodissociation of H$_2$O is the source of OH in higher layers. In this section, we look at the implications for recent and potentially forthcoming SO$_2$ observations.

\subsection{The importance of the stellar flux}
Currently, the UV flux of the host star is one of the major uncertainties in photochemical modeling. Most research so far has taken an approach that combines observations and theoretical models to reconstruct the stellar UV field. \bb{Based on the conclusions from \cite{tsai_photochemically_2023}}, it was logical to assume that stellar UV flux below 200 nm is a necessary condition to observe SO$_2$, but this paper suggests that we can relax this assumption. Quiescent stars with low UV fluxes are an equally promising target when it comes to the detection of photochemistry.  

One may argue that the situation of all flux removed below a certain wavelength is an unphysical situation, but the question is whether this is in fact true. In current photochemical models, the upper boundary is where the atmosphere ends, and it is assumed that there is no UV absorption above the upper boundary. However, models and observations of atmospheric escape suggest that strong XUV absorption in the exosphere is necessary to drive the hydrodynamic escape of the atmosphere \citep{oklopcic_new_2018, Spake_Helium_2018, Allart_spectrally_2018, macleod_stellar_2022, Linssen_expanding_2023, Nail_effects_2023, Zhang_giant_2023, Owen_mapping_2024}. Most of the absorption will be due to atomic hydrogen and therefore occurs below the lyman-$\alpha$ line, suggesting this effect will be limited to the upper atmosphere. To our knowledge, photochemical models that filter the input stellar spectrum through the exosphere have so far not been considered, and we suggest it as a direction for future research. 

The opposite situation can also occur, in which there is significant stellar flux in the UV but lower flux in the NUV. This is the situation in M stars that have significant chromospheric activity, but due to their low effective temperatures have low photospheric flux at higher wavelengths. We suggest targeting close-in gas giants around these stars to investigate the presence of SO$_2$. We expect SO$_2$ to be produced only in higher layers of the atmosphere in these planets, which should lead to a lower visibility of SO$_2$ in the spectrum. 

\subsection{Temperature-dependent UV cross sections}
Laboratory measurements of the UV cross sections of molecules are usually obtained at room temperature\bb{, and there is a need for the expansion of these measurements to the higher temperature ranges such as can be found in hot gas giants \citep{Fortney_need_2019, Venot_VUV_2018}. We use room temperature cross sections for the modeling performed in this paper}. It was suggested by \cite{Heays_photodissociation_2017} that a temperature increase of only a few hundred Kelvin should not impact the UV cross sections majorly for many molecules. For those molecules with transitions between excited vibrational states, however, the impact on the cross sections can in fact be significant. Temperature dependent cross sections for SH are available from the ExoMol project \citep{Gorman_exomol_SH_2019}, who found that the UV cross section increases significantly as the temperature is increased. This occurs in the important 200-350 nm window, as well as above 350 nm. This potentially leads to increased SH photodissociation and thus an increased rate of photochemically induced H$_2$S destruction, which further underscores the importance of the results of this paper for the production of SO$_2$. These cross sections have been implemented in VULCAN, and a number of preliminary tests can be found in \cite{tsai_comparative_2021}. \bb{In addition, temperature dependent cross sections are also available for SO$_2$, although they have only been taken for one temperature so far \citep{Fateev_experimental_2021}.} We leave an investigation of the effect of temperature dependent UV cross sections to a future study.   

\subsection{The case of WASP-107b} \label{The case of wasp-107b}
At the time of writing this paper, SO$_2$ has been detected in the atmospheres of WASP-39b \citep{rustamkulov_early_2023} and WASP-107b \citep{dyrek_so_2_2023}. In light of the results of this paper, the WASP-107b detection is particularly interesting, where alongside SO$_2$ also silicate clouds were detected. 

The pressure level at which the clouds were detected was still relatively poorly constrained. Using two different retrieval codes, the pressure level of the clouds was found to be either 10$^{-5}$ bar or 10$^{-3}$ bar. Additionally, no constraints were placed on the longitude of the clouds, due to the 1D nature of the retrieval framework. Silicate clouds are significant absorbers of UV radiation, so the precise location of the clouds matters significantly for the photochemistry that can occur in an exoplanet. Unfortunately, full cloud coverage obscures the visibility of chemical species below the clouds. However, in the case of WASP-107b, cloud coverage was relatively thin, and atmospheric layers below the clouds contributed partially to the transit spectrum, although this was not the case for the SO$_2$ absorption, which originated from above the clouds. 

In this paper, we found that NUV radiation leads to detectable quantities of photochemically produced SO$_2$ at pressures as high as 10$^{-4}$ bar in a cloud-free model. On a planet with sufficiently low atmospheric temperature, such as WASP-107b, silicate cloud formation can take place on the dayside of the planet, which would be able to partially insulate deeper layers of the atmosphere from the effects of photochemistry. Further complexity is brought about by the effect of global chemical transport around the equator, which can transport the products of photochemistry from the dayside to the nightside, as has been investigated by e.g. \cite{Baeyens_Photodissociation_2023} and \cite{Tsai_day_night_2023}. It is still an open question to what extent partial cloud coverage affects the formation of SO$_2$ below the clouds. We suggest an investigation of the competing effects of clouds and photochemistry as the direction of a potential future study with the available 2D photochemical kinetics codes, such as for example \citep{Tsai_Global_2024}. Planets with thin cloud coverage such as WASP-107b may present an opportunity to constrain the efficiency of photochemistry below the cloud layer.

\section{Summary and conclusion} \label{Conclusion}

 We studied the chemical pathways leading to SO$_2$, as well as the dependency on the stellar flux field. We arrived at the following conclusions:

 \begin{enumerate}
     \item Stellar flux between 300 and 350 nm leads to strong production of SO$_2$ in the absence of stellar flux below 300 nm. 
     \item \bb{NUV driven photochemical dissociation of SH and S$_2$ leads to the production of atomic S and H, which subsequently leads to the formation of SO$_2$.}
     \item The pathway to SO$_2$ is dependent on the atmospheric layer where it is formed. \bb{At pressures around $10^{-4}$ bar}, it depends on the hydrogen abstraction of H$_2$O. \bb{At pressures around $10^{-6}$ bar}, it depends on the photochemical dissociation of H$_2$O.
     \item A lower surface gravity elevates the mixing ratio of SO$_2$ and therefore increases its visibility in the transmission spectrum. This is due to the increased scale height of the atmosphere and stellar radiation being absorbed higher up in the atmosphere.
      \item Hydrogen abstraction of SH plays a role in the pathway to SO$_2$ \bb{at pressures around $10^{-6}$ bar} at temperatures below 800 K.
 \end{enumerate}

\begin{acknowledgements}

S.-M.T. acknowledges support from NASA Exobiology Grant No. 80NSSC20K1437 and the University of California at Riverside. L.D.\ acknowledges funding from the KU Leuven Interdisciplinary Grant (IDN/19/028), the European Union H2020-MSCA-ITN-2019 under Grant no. 860470 (CHAMELEON) and the FWO research grant G086217N. 
\end{acknowledgements}

\bibliographystyle{aa}
\bibliography{refs}

\begin{thebibliography}{45}
\expandafter\ifx\csname natexlab\endcsname\relax\def\natexlab#1{#1}\fi

\bibitem[{{Allart} {et~al.}(2018){Allart}, {Bourrier}, {Lovis}, {Ehrenreich}, {Spake}, {Wyttenbach}, {Pino}, {Pepe}, {Sing}, \& {Lecavelier des Etangs}}]{Allart_spectrally_2018}
{Allart}, R., {Bourrier}, V., {Lovis}, C., {et~al.} 2018, Science, 362, 1384

\bibitem[{{Asplund} {et~al.}(2009){Asplund}, {Grevesse}, {Sauval}, \& {Scott}}]{asplund_solar_abundances_2009}
{Asplund}, M., {Grevesse}, N., {Sauval}, A.~J., \& {Scott}, P. 2009, \araa, 47, 481

\bibitem[{{Baeyens, Robin} {et~al.}(2024){Baeyens, Robin}, {Désert, Jean-Michel}, {Petrignani, Annemieke}, {Carone, Ludmila}, \& {Schneider, Aaron David}}]{Baeyens_Photodissociation_2023}
{Baeyens, Robin}, {Désert, Jean-Michel}, {Petrignani, Annemieke}, {Carone, Ludmila}, \& {Schneider, Aaron David}. 2024, A\&A, 686, A24

\bibitem[{{Boley}(2009)}]{Boley_two_2009}
{Boley}, A.~C. 2009, \apjl, 695, L53

\bibitem[{{Chubb} {et~al.}(2024){Chubb}, {Robert}, {Sousa-Silva}, {Yurchenko}, {Allard}, {Boudon}, {Buldyreva}, {Bultel}, {Coustenis}, {Foltynowicz}, {Gordon}, {Hargreaves}, {Helling}, {Hill}, {Rafn Hrodmarsson}, {Karman}, {Lecoq-Molinos}, {Migliorini}, {Rey}, {Richard}, {Sadiek}, {Schmidt}, {Sokolov}, {Stefani}, {Tennyson}, {Venot}, {Wright}, {Arenales-Lope}, {Barstow}, {Bocchieri}, {Carrasco}, {Dubey}, {Egorov}, {Garc{\'\i}a Mu{\~n}oz}, {Ehsan}, {Gharib-Nezhad}, {Gkouvelis}, {Gr{\"u}bel}, {Irwin}, {Kn{\'\i}{\v{z}}ek}, {Lewis}, {Lodge}, {Ma}, {Martins}, {Molaverdikhani}, {Morello}, {Nikitin}, {Panek}, {Rengel}, {Rinaldi}, {Skinner}, {Tinetti}, {van Kempen}, {Yang}, \& {Zingales}}]{chubb_data_2024}
{Chubb}, K.~L., {Robert}, S., {Sousa-Silva}, C., {et~al.} 2024, arXiv e-prints, arXiv:2404.02188

\bibitem[{Chubb {et~al.}(2021)Chubb, Rocchetto, Yurchenko, Min, Waldmann, Barstow, Mollière, Al-Refaie, Phillips, \& Tennyson}]{chubb_exomolop_2021}
Chubb, K.~L., Rocchetto, M., Yurchenko, S.~N., {et~al.} 2021, A\&A, 646, A21

\bibitem[{{Dyrek} {et~al.}(2024){Dyrek}, {Min}, {Decin}, {Bouwman}, {Crouzet}, {Molli{\`e}re}, {Lagage}, {Konings}, {Tremblin}, {G{\"u}del}, {Pye}, {Waters}, {Henning}, {Vandenbussche}, {Ardevol Martinez}, {Argyriou}, {Ducrot}, {Heinke}, {van Looveren}, {Absil}, {Barrado}, {Baudoz}, {Boccaletti}, {Cossou}, {Coulais}, {Edwards}, {Gastaud}, {Glasse}, {Glauser}, {Greene}, {Kendrew}, {Krause}, {Lahuis}, {Mueller}, {Olofsson}, {Patapis}, {Rouan}, {Royer}, {Scheithauer}, {Waldmann}, {Whiteford}, {Colina}, {van Dishoeck}, {{\"O}stlin}, {Ray}, \& {Wright}}]{dyrek_so_2_2023}
{Dyrek}, A., {Min}, M., {Decin}, L., {et~al.} 2024, \nat, 625, 51

\bibitem[{{Fateev}(2021)}]{Fateev_experimental_2021}
{Fateev}, A. 2021, J. Quant. Spectrosc. Radiat. Transf.

\bibitem[{{Fortney} {et~al.}(2019){Fortney}, {Robinson}, {Domagal-Goldman}, {Genio}, {Gordon}, {Gharib-Nezhad}, {Lewis}, {Sousa-Silva}, {Airapetian}, {Drouin}, {Hargreaves}, {Huang}, {Karman}, {Ramirez}, {Rieker}, {Tennyson}, {Wordsworth}, {Yurchenko}, {Johnson}, {Lee}, {Marley}, {Dong}, {Kane}, {L{\'o}pez-Morales}, {Fauchez}, {Lee}, {Sung}, {Haghighipour}, {Horst}, {Gao}, {Kao}, {Dressing}, {Lupu}, {Savin}, {Fleury}, {Venot}, {Ascenzi}, {Milam}, {Linnartz}, {Gudipati}, {Gronoff}, {Salama}, {Gavilan}, {Bouwman}, {Turbet}, {Benilan}, {Henderson}, {Batalha}, {Jensen-Clem}, {Lyons}, {Freedman}, {Schwieterman}, {Goyal}, {Mancini}, {Irwin}, {Desert}, {Molaverdikhani}, {Gizis}, {Taylor}, {Lothringer}, {Pierrehumbert}, {Zellem}, {Batalha}, {Rugheimer}, {Lustig-Yaeger}, {Hu}, {Kempton}, {Arney}, {Line}, {Alam}, {Moses}, {Iro}, {Kreidberg}, {Blecic}, {Louden}, {Molli{\`e}re}, {Stevenson}, {Swain}, {Bott}, {Madhusudhan}, {Krissansen-Totton}, {Deming}, {Kitiashvili}, {Shkolnik}, {Rustamkulov}, {Rogers}, \&
  {Close}}]{Fortney_need_2019}
{Fortney}, J., {Robinson}, T.~D., {Domagal-Goldman}, S., {et~al.} 2019, Astro2020: Decadal Survey on Astronomy and Astrophysics, 2020, 146

\bibitem[{{Goody} {et~al.}(1989){Goody}, {West}, {Chen}, \& {Crisp}}]{goody_correlated-k_1989}
{Goody}, R., {West}, R., {Chen}, L., \& {Crisp}, D. 1989, \jqsrt, 43, 191

\bibitem[{{Gordon} {et~al.}(2017){Gordon}, {Rothman}, {Hill}, {Kochanov}, {Tan}, {Bernath}, {Birk}, {Boudon}, {Campargue}, {Chance}, {Drouin}, {Flaud}, {Gamache}, {Hodges}, {Jacquemart}, {Perevalov}, {Perrin}, {Shine}, {Smith}, {Tennyson}, {Toon}, {Tran}, {Tyuterev}, {Barbe}, {Cs{\'a}sz{\'a}r}, {Devi}, {Furtenbacher}, {Harrison}, {Hartmann}, {Jolly}, {Johnson}, {Karman}, {Kleiner}, {Kyuberis}, {Loos}, {Lyulin}, {Massie}, {Mikhailenko}, {Moazzen-Ahmadi}, {M{\"u}ller}, {Naumenko}, {Nikitin}, {Polyansky}, {Rey}, {Rotger}, {Sharpe}, {Sung}, {Starikova}, {Tashkun}, {Auwera}, {Wagner}, {Wilzewski}, {Wcis{\l}o}, {Yu}, \& {Zak}}]{Gordon_HITRAN_2017}
{Gordon}, I.~E., {Rothman}, L.~S., {Hill}, C., {et~al.} 2017, \jqsrt, 203, 3

\bibitem[{{Gorman} {et~al.}(2019){Gorman}, {Yurchenko}, \& {Tennyson}}]{Gorman_exomol_SH_2019}
{Gorman}, M.~N., {Yurchenko}, S.~N., \& {Tennyson}, J. 2019, \mnras, 490, 1652

\bibitem[{Guillot(2010)}]{guillot_radiative_2010}
Guillot, T. 2010, A\&A, 520, A27

\bibitem[{{Heays} {et~al.}(2017){Heays}, {Bosman}, \& {van Dishoeck}}]{Heays_photodissociation_2017}
{Heays}, A.~N., {Bosman}, A.~D., \& {van Dishoeck}, E.~F. 2017, \aap, 602, A105

\bibitem[{{Hobbs} {et~al.}(2021){Hobbs}, {Rimmer}, {Shorttle}, \& {Madhusudhan}}]{Hobbs_sulfur_2021}
{Hobbs}, R., {Rimmer}, P.~B., {Shorttle}, O., \& {Madhusudhan}, N. 2021, \mnras, 506, 3186

\bibitem[{{Lacis} \& {Oinas}(1991)}]{lacis_description_1991}
{Lacis}, A.~A. \& {Oinas}, V. 1991, \jgr, 96, 9027

\bibitem[{{Linssen} \& {Oklop{\v{c}}i{\'c}}(2023)}]{Linssen_expanding_2023}
{Linssen}, D.~C. \& {Oklop{\v{c}}i{\'c}}, A. 2023, \aap, 675, A193

\bibitem[{MacLeod \& Oklopčić(2022)}]{macleod_stellar_2022}
MacLeod, M. \& Oklopčić, A. 2022, ApJ, 926, 226

\bibitem[{{Madhusudhan}(2019)}]{Madhusudhan_Exoplanetary_2019}
{Madhusudhan}, N. 2019, \araa, 57, 617

\bibitem[{{Min, Michiel} {et~al.}(2020){Min, Michiel}, {Ormel, Chris W.}, {Chubb, Katy}, {Helling, Christiane}, \& {Kawashima, Yui}}]{min_arcis_2020}
{Min, Michiel}, {Ormel, Chris W.}, {Chubb, Katy}, {Helling, Christiane}, \& {Kawashima, Yui}. 2020, A\&A, 642, A28

\bibitem[{{Moses} {et~al.}(2011){Moses}, {Visscher}, {Fortney}, {Showman}, {Lewis}, {Griffith}, {Klippenstein}, {Shabram}, {Friedson}, {Marley}, \& {Freedman}}]{Moses_Disequilibrium_2011}
{Moses}, J.~I., {Visscher}, C., {Fortney}, J.~J., {et~al.} 2011, \apj, 737, 15

\bibitem[{{Nail, F.} {et~al.}(2024){Nail, F.}, {Oklopčić, A.}, \& {MacLeod, M.}}]{Nail_effects_2023}
{Nail, F.}, {Oklopčić, A.}, \& {MacLeod, M.} 2024, A\&A, 684, A20

\bibitem[{{{\"O}berg} {et~al.}(2011){{\"O}berg}, {Murray-Clay}, \& {Bergin}}]{Oberg_effects_2011}
{{\"O}berg}, K.~I., {Murray-Clay}, R., \& {Bergin}, E.~A. 2011, \apjl, 743, L16

\bibitem[{Oklopčić \& Hirata(2018)}]{oklopcic_new_2018}
Oklopčić, A. \& Hirata, C.~M. 2018, ApJ, 855, L11

\bibitem[{Ormel \& Min(2019)}]{ormel_arcis_2019}
Ormel, C.~W. \& Min, M. 2019, A\&A, 622, A121

\bibitem[{{Owen} \& {Schlichting}(2024)}]{Owen_mapping_2024}
{Owen}, J.~E. \& {Schlichting}, H.~E. 2024, \mnras, 528, 1615

\bibitem[{{Polman, J.} {et~al.}(2023){Polman, J.}, {Waters, L. B. F. M.}, {Min, M.}, {Miguel, Y.}, \& {Khorshid, N.}}]{polman_h2s_2023}
{Polman, J.}, {Waters, L. B. F. M.}, {Min, M.}, {Miguel, Y.}, \& {Khorshid, N.} 2023, A\&A, 670, A161

\bibitem[{{Rafikov}(2005)}]{Rafikov_Giant_2005}
{Rafikov}, R.~R. 2005, \apjl, 621, L69

\bibitem[{{Rimmer} {et~al.}(2021){Rimmer}, {Jordan}, {Constantinou}, {Woitke}, {Shorttle}, {Hobbs}, \& {Paschodimas}}]{Rimmer_Hydroxide_2021}
{Rimmer}, P.~B., {Jordan}, S., {Constantinou}, T., {et~al.} 2021, PSJ, 2, 133

\bibitem[{{Rothman} {et~al.}(2010){Rothman}, {Gordon}, {Barber}, {Dothe}, {Gamache}, {Goldman}, {Perevalov}, {Tashkun}, \& {Tennyson}}]{Rothman_HITEMP_2010}
{Rothman}, L.~S., {Gordon}, I.~E., {Barber}, R.~J., {et~al.} 2010, \jqsrt, 111, 2139

\bibitem[{Rustamkulov {et~al.}(2023)Rustamkulov, Sing, Mukherjee, May, Kirk, Schlawin, Line, Piaulet, Carter, Batalha, Goyal, López-Morales, Lothringer, MacDonald, Moran, Stevenson, Wakeford, Espinoza, Bean, Batalha, Benneke, Berta-Thompson, Crossfield, Gao, Kreidberg, Powell, Cubillos, Gibson, Leconte, Molaverdikhani, Nikolov, Parmentier, Roy, Taylor, Turner, Wheatley, Aggarwal, Ahrer, Alam, Alderson, Allen, Banerjee, Barat, Barrado, Barstow, Bell, Blecic, Brande, Casewell, Changeat, Chubb, Crouzet, Daylan, Decin, Désert, Mikal-Evans, Feinstein, Flagg, Fortney, Harrington, Heng, Hong, Hu, Iro, Kataria, Kempton, Krick, Lendl, Lillo-Box, Louca, Lustig-Yaeger, Mancini, Mansfield, Mayne, Miguel, Morello, Ohno, Palle, Petit Dit De La~Roche, Rackham, Radica, Ramos-Rosado, Redfield, Rogers, Shkolnik, Southworth, Teske, Tremblin, Tucker, Venot, Waalkes, Welbanks, Zhang, \& Zieba}]{rustamkulov_early_2023}
Rustamkulov, Z., Sing, D.~K., Mukherjee, S., {et~al.} 2023, Nature, 614, 659

\bibitem[{{Southworth}(2010)}]{southworth_homogeneous_2010}
{Southworth}, J. 2010, \mnras, 408, 1689

\bibitem[{{Spake} {et~al.}(2018){Spake}, {Sing}, {Evans}, {Oklop{\v{c}}i{\'c}}, {}, {Bourrier}, {Kreidberg}, {Rackham}, {Irwin}, {Ehrenreich}, {Wyttenbach}, {Wakeford}, {Zhou}, {Chubb}, {Nikolov}, {Goyal}, {Henry}, {Williamson}, {Blumenthal}, {Anderson}, {Hellier}, {Charbonneau}, {Udry}, \& {Madhusudhan}}]{Spake_Helium_2018}
{Spake}, J.~J., {Sing}, D.~K., {Evans}, T.~M., {et~al.} 2018, \nat, 557, 68

\bibitem[{{Tennyson} {et~al.}(2016){Tennyson}, {Yurchenko}, {Al-Refaie}, {Barton}, {Chubb}, {Coles}, {Diamantopoulou}, {Gorman}, {Hill}, {Lam}, {Lodi}, {McKemmish}, {Na}, {Owens}, {Polyansky}, {Rivlin}, {Sousa-Silva}, {Underwood}, {Yachmenev}, \& {Zak}}]{Tennyson_ExoMol_2016}
{Tennyson}, J., {Yurchenko}, S.~N., {Al-Refaie}, A.~F., {et~al.} 2016, J. Mol. Spectrosc., 327, 73

\bibitem[{Tsai {et~al.}(2023)Tsai, Lee, Powell, Gao, Zhang, Moses, Hébrard, Venot, Parmentier, Jordan, Hu, Alam, Alderson, Batalha, Bean, Benneke, Bierson, Brady, Carone, Carter, Chubb, Inglis, Leconte, Line, López-Morales, Miguel, Molaverdikhani, Rustamkulov, Sing, Stevenson, Wakeford, Yang, Aggarwal, Baeyens, Barat, De~Val-Borro, Daylan, Fortney, France, Goyal, Grant, Kirk, Kreidberg, Louca, Moran, Mukherjee, Nasedkin, Ohno, Rackham, Redfield, Taylor, Tremblin, Visscher, Wallack, Welbanks, Youngblood, Ahrer, Batalha, Behr, Berta-Thompson, Blecic, Casewell, Crossfield, Crouzet, Cubillos, Decin, Désert, Feinstein, Gibson, Harrington, Heng, Henning, Kempton, Krick, Lagage, Lendl, Lothringer, Mansfield, Mayne, Mikal-Evans, Palle, Schlawin, Shorttle, Wheatley, \& Yurchenko}]{tsai_photochemically_2023}
Tsai, S.-M., Lee, E. K.~H., Powell, D., {et~al.} 2023, Nature, 617, 483

\bibitem[{Tsai {et~al.}(2017)Tsai, Lyons, Grosheintz, Rimmer, Kitzmann, \& Heng}]{tsai_vulcan_2017}
Tsai, S.-M., Lyons, J.~R., Grosheintz, L., {et~al.} 2017, ApJ Supplement Series, 228, 20

\bibitem[{Tsai {et~al.}(2021)Tsai, Malik, Kitzmann, Lyons, Fateev, Lee, \& Heng}]{tsai_comparative_2021}
Tsai, S.-M., Malik, M., Kitzmann, D., {et~al.} 2021, ApJ, 923, 264

\bibitem[{{Tsai} {et~al.}(2023){Tsai}, {Moses}, {Powell}, \& {Lee}}]{Tsai_day_night_2023}
{Tsai}, S.-M., {Moses}, J.~I., {Powell}, D., \& {Lee}, E. K.~H. 2023, \apjl, 959, L30

\bibitem[{{Tsai} {et~al.}(2024){Tsai}, {Parmentier}, {Mendon{\c{c}}a}, {Tan}, {Deitrick}, {Hammond}, {Savel}, {Zhang}, {Pierrehumbert}, \& {Schwieterman}}]{Tsai_Global_2024}
{Tsai}, S.-M., {Parmentier}, V., {Mendon{\c{c}}a}, J.~M., {et~al.} 2024, \apj, 963, 41

\bibitem[{{Turrini} {et~al.}(2021){Turrini}, {Schisano}, {Fonte}, {Molinari}, {Politi}, {Fedele}, {Pani{\'c}}, {Kama}, {Changeat}, \& {Tinetti}}]{Turrini_Tracing_2021}
{Turrini}, D., {Schisano}, E., {Fonte}, S., {et~al.} 2021, \apj, 909, 40

\bibitem[{{Venot} {et~al.}(2018){Venot}, {Bénilan, Y.}, {Fray, N.}, {Gazeau, M.-C.}, {Lefèvre, F.}, {Es-sebbar, Et.}, {Hébrard, E.}, {Schwell, M.}, {Bahrini, C.}, {Montmessin, F.}, {Lefèvre, M.}, \& {Waldmann, I. P.}}]{Venot_VUV_2018}
{Venot}, {Bénilan, Y.}, {Fray, N.}, {et~al.} 2018, A\&A, 609, A34

\bibitem[{{Venot} {et~al.}(2012){Venot}, {H{\'e}brard}, {Ag{\'u}ndez}, {Dobrijevic}, {Selsis}, {Hersant}, {Iro}, \& {Bounaceur}}]{Venot_chemical_2012}
{Venot}, O., {H{\'e}brard}, E., {Ag{\'u}ndez}, M., {et~al.} 2012, \aap, 546, A43

\bibitem[{{Yurchenko} {et~al.}(2018){Yurchenko}, {Al-Refaie}, \& {Tennyson}}]{Yurchenko_2018}
{Yurchenko}, S.~N., {Al-Refaie}, A.~F., \& {Tennyson}, J. 2018, \aap, 614, A131

\bibitem[{{Zahnle} {et~al.}(2009){Zahnle}, {Marley}, {Freedman}, {Lodders}, \& {Fortney}}]{Zahnle_atmospheric_sulfur_2009}
{Zahnle}, K., {Marley}, M.~S., {Freedman}, R.~S., {Lodders}, K., \& {Fortney}, J.~J. 2009, \apjl, 701, L20

\bibitem[{{Zhang} {et~al.}(2023){Zhang}, {Morley}, {Gully-Santiago}, {MacLeod}, {Oklop{\v{c}}i{\'c}}, {Luna}, {Tran}, {Ninan}, {Mahadevan}, {Krolikowski}, {Cochran}, {Bowler}, {Endl}, {Stef{\'a}nsson}, {Tofflemire}, {Vanderburg}, \& {Zeimann}}]{Zhang_giant_2023}
{Zhang}, Z., {Morley}, C.~V., {Gully-Santiago}, M., {et~al.} 2023, Science Advances, 9, eadf8736

\end{thebibliography}

\begin{appendix}
\section{Test of chemical equilibrium}
\begin{figure}[h]
    \centering
    \includegraphics[width=\linewidth]{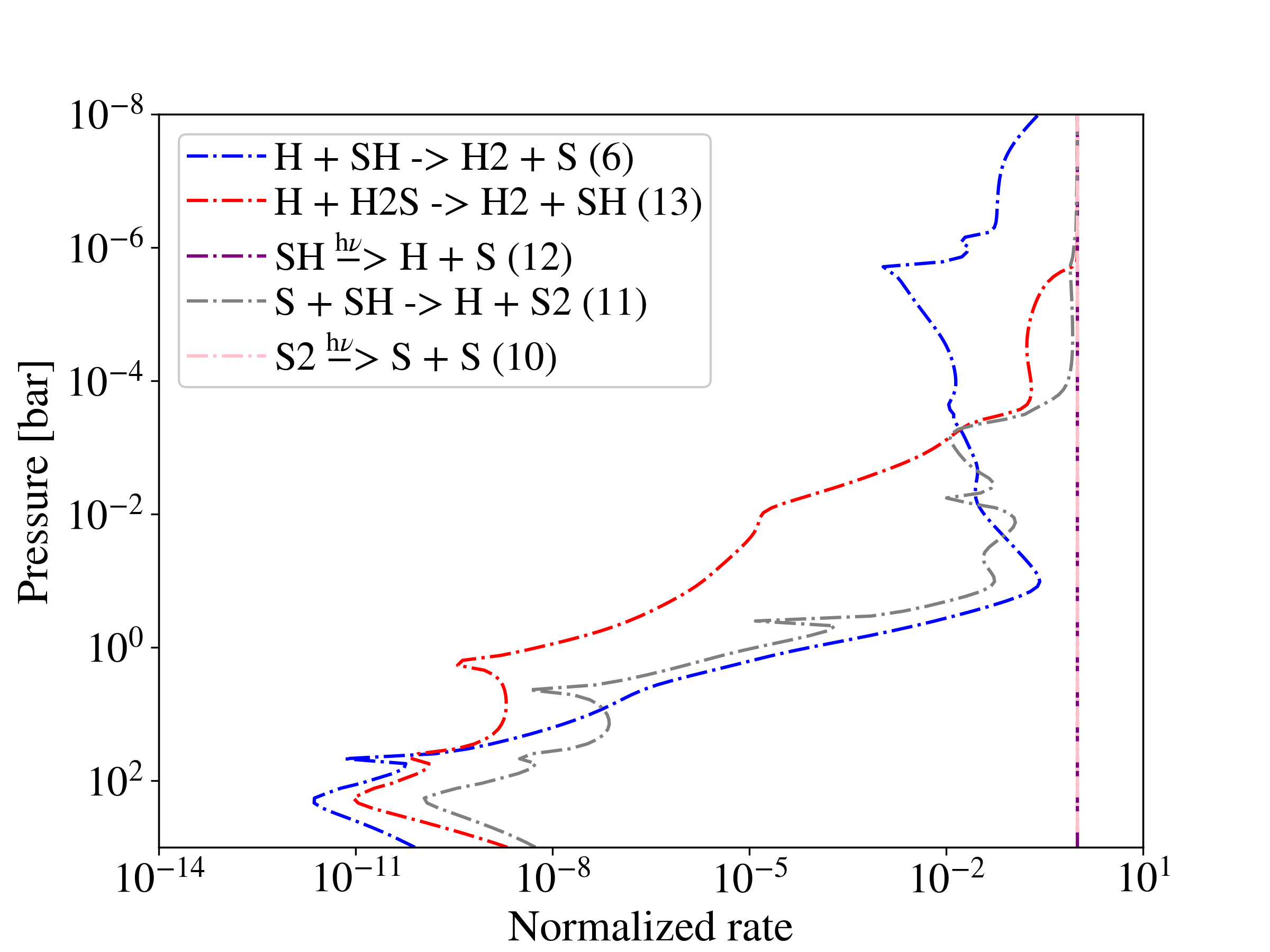}
    \caption{\bb{Normalized reaction rates (abs(forward reaction rate - reverse reaction rate)/(forward reaction rate)) of various chemical reactions in the fiducial model. Normalized reaction rates become comparatively small in the bottom regions of the atmosphere, indicating these regions approach chemical equilibrium.}}
    \label{fig:Normalized_rates}
\end{figure}
\end{appendix}

\end{document}